\documentclass[pdflatex,sn-mathphys-num]{sn-jnl}

\usepackage{graphicx}%
\usepackage{multirow}%
\usepackage{amsmath,amssymb,amsfonts}%
\usepackage{amsthm}%
\usepackage{mathrsfs}%
\usepackage[title]{appendix}%
\usepackage{xcolor}%
\usepackage{textcomp}%
\usepackage{manyfoot}%
\usepackage{booktabs}%
\usepackage{algorithm}%
\usepackage{algorithmicx}%
\usepackage{algpseudocode}%
\usepackage{listings}%
\usepackage{array}
\usepackage{caption}
\usepackage{threeparttable}
\newcolumntype{P}[1]{>{\centering\arraybackslash}p{#1}}
\newcolumntype{M}[1]{>{\centering\arraybackslash}m{#1}}

\theoremstyle{thmstyleone}%
\theoremstyle{thmstyletwo}%

\theoremstyle{thmstylethree}%

\begin{document}

\title[Article Title]{Regression Testing Optimization for ROS-based Autonomous Systems: A Comprehensive Review of Techniques}


\author[1]{\fnm{Yupeng} \sur{Jiang}}\email{yupeng.jiang@hdr.mq.edu.au}
\equalcont{These authors contributed equally to this work.}

\author[2]{\fnm{Shuaiyi} \sur{Sun}}\email{D23092100498@cityu.edu.mo}
\equalcont{These authors contributed equally to this work.}

\author*[1]{\fnm{Xi} \sur{Zheng}}\email{james.zheng@mq.edu.au}

\affil[1]{\orgname{Macquarie University}, \orgaddress{\city{Sydney}, \country{Australia}}}

\affil[2]{\orgname{City University of Macau}, \orgaddress{\city{Macau}, \country{China}}}





\abstract{Regression testing plays a critical role in maintaining software reliability, particularly for ROS-based autonomous systems (ROSAS), which frequently undergo continuous integration and iterative development. However, conventional regression testing techniques face significant challenges when applied to autonomous systems due to their dynamic and non-deterministic behaviors, complex multi-modal sensor data, asynchronous distributed architectures, and stringent safety and real-time constraints. Although numerous studies have explored test optimization in traditional software contexts, regression testing optimization specifically for ROSAS remains largely unexplored. To address this gap, we present the first comprehensive survey systematically reviewing regression testing optimization techniques tailored for ROSAS. We analyze and categorize 122 representative studies into regression test case prioritization, minimization, and selection methods. A structured taxonomy is introduced to clearly illustrate their applicability and limitations within ROSAS contexts. Furthermore, we highlight major challenges specific to regression testing for ROSAS, including effectively prioritizing tests in response to frequent system modifications, efficiently minimizing redundant tests, and difficulty in accurately selecting impacted test cases. Finally, we propose research insights and identify promising future directions, such as leveraging frame-to-vector coverage metrics, multi-source foundation models, and neurosymbolic reasoning to enhance regression testing efficiency and effectiveness. This survey provides a foundational reference and practical roadmap for advancing the state-of-the-art in regression testing optimization for ROSAS.}

\keywords{Robot Operating System, test prioritization, test reduction, test selection, regression testing}



\maketitle

\section{Introduction}\label{sec1}

\begin{figure*}
	\setlength{\belowcaptionskip}{-0.3cm}
	\centering
	\includegraphics[width=\textwidth]{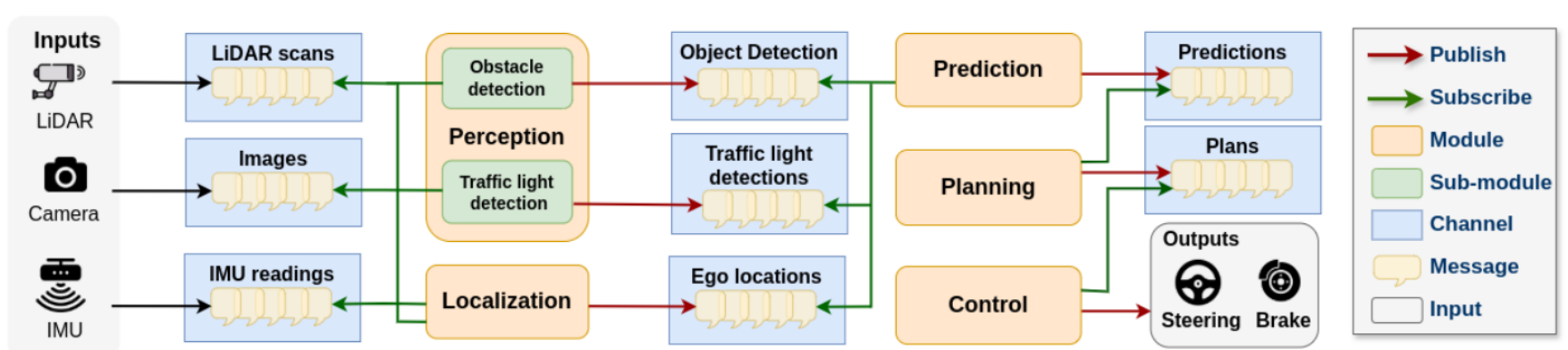}
        \vspace{3pt}
        \captionsetup{font=small}
	\caption{A typical architecture of a ROS-based autonomous system \cite{deng2022scenario}.}
	\label{ROSAS} 
\end{figure*}

Autonomous systems, such as self-driving cars (Waymo \cite{Waymo}, Tesla Autopilot \cite{Autopilot}, Uber AV \cite{Uber}), aerial drones (Wing \cite{Wing}, Zipline \cite{Zipline}, Percepto \cite{Percepto}), and industrial robots (Unitree \cite{Unitree}, Ubtech \cite{UBTECH}, Boston Dynamics \cite{BostonDynamics}, Da Vinci \cite{DaVinci}), are increasingly becoming integral to various safety-critical applications. Many of these systems are built on top of the Robot Operating System (ROS) \cite{quigley2009ros}, which provides a flexible and modular middleware framework to support distributed processing, multi-sensor integration, and dynamic task execution. The architecture of a typical ROS-based autonomous system (ROSAS) is illustrated in Figure~\ref{ROSAS}. This system is designed to process inputs from multiple sensor modalities, including LiDAR, cameras, and Inertial Measurement Units (IMUs), to achieve comprehensive environmental perception and autonomous decision-making. As ROSAS continue to evolve rapidly in complexity and scale, ensuring their reliability and safety through rigorous testing becomes increasingly essential. The current body of work on ROSAS testing encompasses a range of domains, including autonomous driving systems (ADS) \cite{deng2022declarative, deng2022scenario}, unmanned aerial vehicles (UAVs) \cite{liang2025garl, schroder2025robustautonomouslandingsystems}, and cyber-physical systems (CPS) \cite{zheng2024testing}.

Unlike traditional software, ROSAS face unique challenges during their development and maintenance life cycles. These systems frequently undergo iterative and incremental updates involving software module refinements, integration of new sensor modalities, configuration adjustments, and behavioral modifications \cite{derler2011modeling}. Each update, whether small or large, has the potential to introduce unintended side effects that may compromise system safety or operational correctness. To mitigate such risks, regression testing plays a crucial role in verifying that new modifications do not negatively impact existing functionalities. Through selective re-execution of relevant test cases, regression testing helps maintain confidence in system stability without incurring the prohibitive cost of exhaustive retesting.

However, conducting regression testing efficiently in ROSAS presents significant challenges \cite{afzal2020study}. These systems often consist of highly distributed and asynchronous components, interact with diverse and dynamic environments, and produce complex multi-modal outputs. Furthermore, regression testing in such contexts must often be performed under tight resource and time constraints, particularly in continuous integration (CI) and deployment workflows. As a result, there is a pressing need for regression testing optimization techniques that can intelligently prioritize, minimize, and select test cases to maximize fault detection while minimizing testing overhead.

Although a substantial body of research exists on test optimization in traditional software engineering, these studies largely focus on general-purpose software systems and do not sufficiently address the specific demands of autonomous systems. Existing methods for test case prioritization, minimization, and selection have limited applicability when faced with the semantic complexity, asynchronous interactions, and real-time requirements characteristic of ROSAS environments. Notably, no existing surveys have systematically explored regression testing optimization techniques from the perspective of autonomous systems, leaving a critical gap in current knowledge.

To address this gap, this paper presents the first comprehensive review dedicated to regression testing optimization for ROSAS. By systematically analyzing relevant studies, we aim to (i) identify major categories of optimization techniques applicable to regression testing in autonomous contexts, (ii) examine their applicability, strengths, and limitations, and (iii) propose research insights and future directions tailored to the unique challenges of ROSAS regression testing.

By bridging the gap between traditional test optimization research and the specific needs of autonomous systems, this survey not only consolidates existing knowledge but also provides a roadmap for future studies aimed at improving the efficiency, effectiveness, and safety assurance of regression testing practices in ROSAS.

\subsection{Inadequacy of Traditional Regression Testing in Autonomous Systems}\label{subsec1}
Despite the maturity of regression testing methodologies in traditional software engineering, there is a noticeable lack of systematic research on regression testing within the domain of autonomous systems. Several fundamental differences between traditional software systems and autonomous robotic systems explain this gap.

First, autonomous systems operate in dynamic, unpredictable, and sensor-driven environments where system behaviors are continuous, non-deterministic, and highly context-dependent. This makes it extremely difficult to define test oracles and to establish clear mappings between software modifications and behavioral changes.

Second, small code changes can cause non-local, cascading impacts across perception, planning, and control modules, complicating change impact analysis. Traditional regression testing relies on the locality of change propagation, an assumption that does not hold in autonomous systems.

Third, managing and versioning test data for autonomous systems is substantially more complex than for traditional software, as test cases often involve large-scale sensor recordings, simulation logs, and real-world driving data, all of which may themselves evolve with software updates.

Fourth, the detection of minor violations is a crucial aspect of autonomous system testing, particularly in domains with high reliability requirements (such as autonomous driving), where even extremely minor errors can potentially lead to severe consequences.

Moreover, the high cost of test execution including hardware-in-the-loop testing, simulation environment maintenance, and real-world validation renders traditional batch-oriented regression testing strategies impractical.

As a result, autonomous system testing demands new paradigms that move beyond classical regression testing models, emphasizing semantic-aware, environment-aware, and adaptive testing strategies to cope with the unique challenges of autonomous behavior validation.

\subsection{Scope of this Survey}
The scope of this survey is explicitly focused on regression testing optimization techniques for ROS-based autonomous systems. Regression testing refers to the re-execution of test cases following system modifications to ensure unchanged functionalities are unaffected. Within this context, we specifically investigate three major categories:
\begin{itemize}
\item \textbf{Test Case Prioritization (TCP):} Techniques aimed at ordering test cases to effectively identify faults early in the modified parts of ROS-based autonomous systems.
\item \textbf{Test Suite Minimization (TSM):} Approaches that systematically eliminate redundant or unnecessary tests to improve the efficiency of regression testing while preserving fault detection capability.
\item \textbf{Test Case Selection (TCS):} Methods designed to select subsets of tests that target only those components or modules impacted by recent system changes, significantly reducing testing overhead.
\end{itemize}

Hybrid and emerging optimization approaches that combine these three categories or incorporate machine learning and neurosymbolic methods are also explored. This focused scope fills a critical gap, as existing reviews have not systematically addressed regression testing optimization within the distinct context of ROS-based autonomous systems, thereby underscoring the novelty and unique contribution of this survey.


\subsection{Relevant Existing Surveys}
To contextualize the unique contributions of this survey, we compare it with representative existing surveys that focus on regression testing optimization techniques. Table~\ref{Existed_Ours_Survey} provides a comparative overview across several dimensions, including the target systems, types of optimization addressed, techniques employed, optimization goals, and distinctive focuses of each survey. This comparison highlights the evolving landscape of test optimization research and positions our survey within this broader context.

\begin{table*}
	\centering
	\caption{Comparison of Our Survey with Representative Existing Surveys}
	\label{Existed_Ours_Survey}
	\begin{threeparttable}
		\renewcommand\arraystretch{1.5}
		\newcolumntype{l}{>{\raggedright}p{0.2\textwidth}}
		\newcolumntype{A}{>{\raggedright}p{0.3\textwidth}}
		\newcolumntype{B}{>{\raggedright\arraybackslash}p{0.2\textwidth}}
		\newcolumntype{C}{>{\raggedright\arraybackslash}p{0.25\textwidth}}
		\newcolumntype{D}{>{\raggedright\arraybackslash}p{0.22\textwidth}}
            \newcolumntype{E}{>{\raggedright\arraybackslash}p{0.23\textwidth}}
		\scalebox{0.6}{
			\begin{tabular}{l A B C D E}
				\toprule
				\textbf{Authors \& Ref} & \textbf{Target System} & \textbf{Optimization Type} & \textbf{Techniques Addressed} & \textbf{Optimization Goal} & \textbf{Insights Proposed} \\
				\midrule
				Yoo and Harman \cite{yoo2012regression} & Traditional software & TCP, TSM, TCS & Search-based, heuristic & Fault detection, coverage, cost & \quad \quad — \\
				\midrule
				Habib et al. \cite{habib2023systematic} & Traditional software & TSM & Search-based (evolutionary, swarm, physics-based) & Cost minimization, size reduction & \quad \quad — \\
				\midrule
				Xiao et al. \cite{xiao2023systematic} & Traditional software & TCP, TCS & Machine learning-based, heuristic & Fault detection, cost & \quad \quad — \\
                \midrule
                Pan et al. \cite{pan2022test} & Traditional software (with CI) & TCP, TCS & Machine learning-based (supervised, reinforcement) & Early fault detection, cost & \quad \quad — \\
                \midrule
                Haas et al. \cite{haas2024optimization} & Industrial systems (Automated + Manual) & TCP, TCS & Impact analysis, Pareto optimization & Early feedback, time reduction & \quad \quad — \\
                \midrule
                Sadri-Moshkenani et al. \cite{sadri2022survey} & Cyber-physical systems & TCP, TCS & Model-based, simulation-based & Coverage, safety, efficiency & \quad \quad — \\
                \midrule
                Ours & ROS-based autonomous systems & TCP, TSM, TCS, new optimization paradigms & Diverse & Safety, tiny violation detection, real-time constraints & Frame-to-Vector coverage, Multi-source foundation models, Neurosymbolic approaches \\
				\bottomrule
		\end{tabular}}
	\end{threeparttable}
\end{table*}

As shown in Table~\ref{Existed_Ours_Survey}, previous surveys have largely concentrated on traditional software systems \cite{yoo2012regression, habib2023systematic, xiao2023systematic, pan2022test}, cyber-physical systems \cite{sadri2022survey}, or industrial testing environments \cite{haas2024optimization}. Yoo and Harman \cite{yoo2012regression} and Habib et al. \cite{habib2023systematic} primarily focus on regression testing optimization methods based on search algorithms and heuristic strategies. Xiao et al. \cite{xiao2023systematic} and Pan et al. \cite{pan2022test} concentrate on the application of machine learning techniques in regression testing optimization. Haas et al. \cite{haas2024optimization} is dedicated to evaluating the practical effectiveness of such techniques in industrial practice. Sadri-Moshkenani et al. \cite{sadri2022survey} specifically analyzes regression testing optimization methods for CPS. They typically classify and review optimization techniques based on goals (e.g., fault detection, coverage maximization, cost reduction), artifacts (e.g., code or model-based), or approaches (e.g., search-based, machine learning-based methods). However, none have specifically addressed the unique challenges associated with ROSAS, such as multi-modal data fusion, dynamic and real-time behaviors, and safety-critical validation. In contrast, our survey is the first to systematically review and synthesize regression testing optimization techniques specifically tailored for ROSAS. By introducing new paradigms such as frame-to-vector framework-guided coverage metrics, multi-source foundation model integration, and neurosymbolic reasoning for tiny violation detection, this work advances the frontier of test optimization research beyond the conventional software and CPS testing domains.

\subsection{Contributions of this Survey}
This survey systematically addresses the research gap concerning regression testing optimization specifically for ROS-based autonomous systems, distinguishing it from existing literature that predominantly addresses traditional software testing contexts. The primary contributions of this paper include:
\begin{itemize}
\item This is the first dedicated review explicitly addressing regression testing optimization techniques tailored to ROS-based autonomous systems, a critical yet significantly under-explored research area.
\item We conducted a rigorous and systematic analysis of 122 representative studies, categorizing them according to regression test case prioritization, regression test suite minimization, and regression test case selection.
\item A clear taxonomy is introduced, systematically organizing and classifying regression testing optimization techniques according to their principles, applicability, and relevance within ROS-based autonomous system testing.
\item Major challenges inherent to regression testing in ROS-based autonomous systems, such as accurately determining impacted tests, efficiently handling test redundancy, and effectively prioritizing tests amid continuous system evolution are thoroughly identified and discussed.
\item Based on identified challenges, we propose insightful research directions, including innovative frame-to-vector coverage metrics, the use of multi-source foundation models for semantic test understanding, and neurosymbolic methods for fine-grained violation detection in regression scenarios.
\end{itemize}

Overall, this survey not only consolidates existing knowledge but also provides a clear, structured research roadmap, guiding future investigations toward more effective, scalable, and reliable regression testing practices for ROS-based autonomous systems. Section~\ref{sec:methodology} presents the research methodology, including the formulation of research questions, the literature search strategy, selection criteria, and the analytical procedures used to ensure rigorous and unbiased coverage of relevant studies. Then we review traditional test optimization techniques in Section~\ref{sec:traditional}, laying the foundational knowledge necessary for understanding their evolution. Section~\ref{sec:ros} focuses on representative and emerging optimization methods that have been adapted or show potential for application in autonomous systems, offering a detailed categorization and analysis of their applicability. In Section~\ref{sec:challenge}, we synthesize the identified challenges and offer research insights along with future research directions, aiming to guide the development of next-generation optimization techniques suitable for ROS-based autonomous systems. Finally, Section~\ref{sec:conclusion} concludes the paper.

\section{Research Methodology}\label{sec:methodology}
To ensure rigor and comprehensiveness in this literature review, we adopted a systematic methodology outlined in \cite{kitchenham_guidelines_2007} that involved clearly defined research questions, structured search strategies, precise inclusion and exclusion criteria, and a thorough categorization and analysis of the selected literature. This section details our methodological approach, providing transparency about the search, selection, and analytical processes, which facilitates reproducibility and ensures the reliability of our findings.

\subsection{Research Questions}
To provide comprehensive insights into test optimization techniques applicable to ROS-based autonomous systems, we defined three main research questions.

\begin{itemize}
\item \textbf{RQ1:} What are the major regression testing optimization techniques applicable to ROS-based autonomous systems?
\item \textbf{RQ2:} What are the limitations and gaps in current optimization methods when applied to these systems?
\item \textbf{RQ3:} What are the promising future directions and emerging techniques for regression testing optimization in ROS-based autonomous systems?
\end{itemize}

These questions guided our literature search, analysis, and subsequent discussions.

\subsection{Literature Search Strategy}
To comprehensively identify relevant literature addressing the research questions, we conducted systematic searches across multiple academic databases, selected for their extensive coverage of high-quality peer-reviewed publications in software engineering and autonomous systems research. Specifically, we included the following databases due to their recognized relevance and comprehensive indexing of scholarly articles in these fields: IEEE Xplore, ACM Digital Library, SpringerLink, Elsevier (ScienceDirect), Wiley Online Library, Curran Associates, IET Digital Library, and World Scientific Publishing.

We employed carefully selected search keywords to ensure thoroughness and precision in capturing studies relevant to test optimization in the context of ROS-based autonomous systems. The search keywords included combinations of key terms such as: ``ROS testing'', ``autonomous systems testing'', ``test optimization'', ``test case prioritization'', ``test suite minimization'', ``test reduction'', ``test case selection'', and ``regression testing''. Initial searches conducted across the listed databases, using combinations of the above keywords, returned a total of 2,100 candidate papers.

\subsection{Selection Criteria}
To systematically determine the most relevant literature, explicit inclusion and exclusion criteria were defined and rigorously applied as follows.

\paragraph{\textbf{Inclusion Criteria}}
Studies were included if they met the following conditions:
\begin{itemize}
\item Clearly addressed topics relevant to test optimization, specifically including: Test Case Prioritization (TCP), Test Suite Minimization (TSM) or Test Reduction, Test Case Selection (TCS), and Regression Testing.
\item Were related directly or indirectly (with potential transferability) to ROS-based or similar autonomous systems.
\item Were published in peer-reviewed academic journals or peer-reviewed conference proceedings.
\item Contained sufficiently detailed methodological descriptions or empirical data to support rigorous analysis.
\end{itemize}

\paragraph{\textbf{Exclusion Criteria}}
Studies were excluded based on the following conditions:
\begin{itemize}
\item Did not explicitly address test optimization methodologies or were unrelated to software or autonomous system testing.
\item Lacked methodological rigor or sufficient details regarding techniques or experimental validation.
\item Were duplicates or significantly overlapped with already included studies.
\item Were non-peer-reviewed publications (e.g., editorials, opinion papers, blogs, books, or informal workshop papers).
\end{itemize}

To minimize the risk of overlooking relevant studies, we further applied the snowballing technique \cite{wohlin2014guidelines}, systematically expanding our literature set by tracing connections from initial relevant papers to additional sources. Specifically, both backward snowballing (reviewing reference lists of selected papers to find prior influential studies) and forward snowballing (identifying newer studies citing the initially selected papers through citation databases) were employed.

Following the application of these inclusion and exclusion criteria, combined with the snowballing method, the final literature set consisted of 122 papers. Among these, 6 were survey or review articles providing foundational or comparative overviews of test optimization techniques, and 116 were technical research papers contributing empirical evidence, methodologies, or experimental results. This structured selection approach ensured comprehensive coverage and rigor in our systematic review.

\subsection{Data Analysis and Categorization}
We systematically categorized the selected 122 papers according to several dimensions.

\paragraph{\textbf{Distribution by Optimization Technique}}
We categorized papers based on the test optimization techniques they addressed. The distribution of selected studies according to the type of optimization technique addressed is shown in Table~\ref{Tech_Distribution}. Some studies simultaneously addressed multiple optimization techniques (e.g., Test Case Prioritization and Test Case Selection), thus resulting in overlap across categories. Consequently, the sum of papers across the three categories (137) exceeds the actual total number of selected studies (122). This overlap highlights the interconnected nature of test optimization techniques, reflecting that researchers often combine multiple methodologies within single studies. Among the three, test case prioritization has received the most attention, followed by test case selection and test suite minimization.

\begin{table}[!ht]\scriptsize
    \centering
    \caption{Distribution of Selected Studies by Optimization Technique}
    \label{Tech_Distribution}
    \begin{tabular*}{0.8\textwidth}{p{6.5cm} c}
    \toprule
    \textbf{Optimization Technique} & \textbf{Number of Papers} \\
    \midrule
    Test Case Prioritization (TCP) & 65 \\
    \midrule
    Test Suite Minimization (TSM) & 29 \\
    \midrule
    Test Case Selection (TCS) & 43 \\
    \midrule
    \textbf{Total (with overlaps)} & \textbf{137} \\
    \midrule
    \textbf{Total unique papers} & \textbf{122} \\
    \bottomrule
    \end{tabular*}
\end{table}

\paragraph{\textbf{Distribution by Publication Year}}
The chronological distribution of the selected papers is summarized in Table~\ref{Year_Distribution}. As shown, there has been significant growth in publications since 2020, highlighting increasing research attention to test optimization methods. This trend suggests that the field is rapidly evolving, with more researchers recognizing the importance of optimizing testing processes to improve software quality and reduce costs.

\begin{table}[!ht]\scriptsize
    \centering
    \caption{Distribution of Selected Studies by Publication Year}
    \label{Year_Distribution}
    \begin{tabular*}{0.75\textwidth}{p{5cm} c}
    \toprule
    \quad \quad \textbf{Year} & \textbf{Number of Papers} \\
    \midrule
    \quad \quad 2012 & 1 \\
    \midrule
    \quad \quad 2013 & 1 \\
    \midrule
    \quad \quad 2017 & 2 \\
    \midrule
    \quad \quad 2018 & 3 \\
    \midrule
    \quad \quad 2019 & 5 \\
    \midrule    
    \quad \quad 2020 & 6 \\
    \midrule
    \quad \quad 2021 & 21 \\
    \midrule
    \quad \quad 2022 & 27 \\
    \midrule
    \quad \quad 2023 & 33 \\
    \midrule
    \quad \quad 2024 & 23 \\
    \midrule
    \quad \quad \textbf{Total} & \textbf{122} \\
    \bottomrule
    \end{tabular*}
\end{table}

\paragraph{\textbf{Distribution by Research Type}}
To provide insights into methodological perspectives, we further classified papers into three categories: traditional software testing, machine learning-based approaches, and hybrid approaches. Table~\ref{Type_Distribution} details this distribution. Most studies have focused on traditional software testing, while research on testing machine learning and hybrid systems remains relatively limited.

\begin{table}[!ht]\scriptsize
    \centering
    \caption{Distribution of Selected Studies by Research Type}
    \label{Type_Distribution}
    \begin{tabular*}{0.75\textwidth}{p{5.5cm} c}
    \toprule
    \quad \quad \textbf{Research Type} & \textbf{Number of Papers} \\
    \midrule
    \quad \quad Traditional Software Testing & 88 \\
    \midrule
    \quad \quad Machine Learning-based & 22 \\
    \midrule
    \quad \quad Hybrid Approaches & 12 \\
    \midrule
    \quad \quad \textbf{Total} & \textbf{122} \\
    \bottomrule
    \end{tabular*}
\end{table}

\paragraph{\textbf{Distribution by Academic Databases}}
The distribution of the selected studies across academic databases is summarized in Table~\ref{Databases_Distribution}. This indicates that the research papers are primarily concentrated in IEEE Xplore, ACM Digital Library, and SpringerLink, which collectively account for 81.1\% of the total. This suggests a high degree of concentration of research outcomes on testing optimization techniques across publishing platforms.

\begin{table}[!ht]\scriptsize
    \centering
    \caption{Distribution of Selected Studies by Academic Databases}
    \label{Databases_Distribution}
    \begin{tabular*}{0.8\textwidth}{p{6cm} c}
    \toprule
    \quad \quad \textbf{Database} & \textbf{Number of Papers} \\
    \midrule
    \quad \quad IEEE Xplore & 45 \\
    \midrule
    \quad \quad ACM Digital Library & 29 \\
    \midrule
    \quad \quad SpringerLink & 25 \\
    \midrule
    \quad \quad Elsevier & 9 \\
    \midrule
    \quad \quad Wiley & 4 \\
    \midrule
    \quad \quad Tech Science & 2 \\
    \midrule
    \quad \quad Curran Associates & 1 \\
    \midrule
    \quad \quad IET & 1 \\
    \midrule
    \quad \quad World Scientific Publishing & 1 \\
    \midrule
    \quad \quad ijcaonline & 1 \\
    \midrule
    \quad \quad JOT & 1 \\
    \midrule
    \quad \quad IOPscience & 1 \\
    \midrule
    \quad \quad igi-global & 1 \\
    \midrule
    \quad \quad JSQ & 1 \\
    \midrule
    \quad \quad \textbf{Total} & \textbf{122} \\
    \bottomrule
    \end{tabular*}
\end{table}

\paragraph{\textbf{Geographical Distribution}}
We also analyzed the geographical origins of the reviewed studies (Table~\ref{Geo_Distribution}). This geographic diversity illustrates widespread global interest in test optimization methods. Different countries and regions have conducted research to varying degrees in this field according to their own scientific research capabilities and research needs. For example, some countries may place greater emphasis on the study and innovation of theoretical methods, while others may focus more on practical applications and technology dissemination. This diversified research pattern helps to promote the comprehensive development and progress of the field.

\begin{table}[!ht]\scriptsize
    \centering
    \caption{Distribution of Selected Studies by Geographical Origins}
    \label{Geo_Distribution}
    \begin{tabular*}{0.84\textwidth}{c c c c c c}
    \toprule
    \textbf{Country} & \textbf{Papers} & \textbf{Country} & \textbf{Papers} & \textbf{Country} & \textbf{Papers} \\
    \midrule
    China & 32 & France & 2 & Sweden & 1 \\
    \midrule
    USA & 14 & Finland & 2 & Switzerland & 5 \\
    \midrule
    Australia & 1 & UK & 4 & Saudi Arabia & 1 \\
    \midrule
    Pakistan & 2 & Canada & 7 & Turkey & 3 \\
    \midrule
    Brazil & 5 & Luxembourg & 5 & Spain & 4 \\
    \midrule
    Germany & 5 & Romania & 1 & Singapore & 1 \\
    \midrule
    Russia & 1 & South Africa & 1 & Tunisia & 1 \\
    \midrule
    Norway & 3 & Vietnam & 1 & Italy & 3 \\
    \midrule
    Portugal & 1 & India & 11 & Bangladesh & 1 \\
    \midrule
    Iran & 2 & Netherlands & 1 & Indonesia & 1 \\
    \bottomrule
    \end{tabular*}
\end{table}

\section{Traditional Regression Testing Optimization Techniques} \label{sec:traditional}
In response to the growing cost and complexity of regression testing in traditional software systems, a wide range of optimization techniques have been proposed and extensively studied. These techniques aim to improve testing efficiency by reducing the number of test cases executed and by accelerating the identification of faults, while maintaining acceptable levels of coverage and fault detection effectiveness.






\subsection{Key Techniques and Representative Studies in Traditional Software Regression Testing} \label{tech_traditional}
Over the past two decades, significant research efforts have been dedicated to developing and evaluating test optimization techniques in traditional software engineering. These studies have laid the foundation for understanding how to improve testing efficiency, effectiveness, and scalability across various types of software systems. Yoo and Harman \cite{yoo2012regression} provided one of the most comprehensive surveys on regression test optimization, systematically categorizing techniques into test suite minimization, test case selection, and test case prioritization. Their work highlighted the inherent trade-offs between maximizing fault detection, minimizing testing costs, and preserving coverage, offering a theoretical framework that continues to guide subsequent research.

Three primary categories of regression testing optimization methods have emerged as the foundation of this research area:

\paragraph{\textbf{Test Case Prioritization (TCP)}}
Test Case Prioritization (TCP)  aims to reorder test cases such that those with the highest likelihood of detecting faults or covering critical functionalities are executed earlier \cite{wong1997study}. Strategies include coverage-based prioritization, where test cases are ranked according to the amount of code (e.g., statements, branches) they exercise \cite{huynh2024segment, HUANG2020110712}, and history-based prioritization, which uses past defect detection information to guide ordering \cite{Rahman2018PrioritizingDT, PRADHAN201986, Garg2024TestCP}.

CCCP \cite{HUANG2020110712} is a regression test case prioritization method based on Code Combinations Coverage (CCC), aiming to enhance fault detection by effectively utilizing code coverage information. Unlike previous techniques that focus solely on individual code units, CCCP considers their combinatorial relationships. It represents test cases as binary tuples and applies combinatorial coverage techniques to prioritize them. This approach provides a comprehensive assessment of test suite effectiveness. CCCP exhibits strong adaptability across various code and test case granularities, with particularly favorable performance observed at the test method level. HSP \cite{MAGALHAES2020110430} is a hybrid strategy that integrates information retrieval and code coverage techniques for test case selection and prioritization in regression testing. Previous code coverage-based methods assess test case effectiveness solely based on code coverage, ignoring the semantic correlation between test cases and software changes. HSP overcomes this limitation by incorporating information retrieval, allowing test case selection to be based on both code coverage and semantic relevance to software changes described in change requests. Experiments show that HSP outperforms either information retrieval or code coverage alone in terms of code coverage and fault detection rate. This indicates that combining semantic and code coverage information can more effectively guide test case prioritization and selection, offering new insights for code coverage-based prioritization techniques. SegTCP \cite{huynh2024segment} innovates on and extends the approach of earlier coverage-based test case prioritization methods. Unlike previous methods that typically focus on a single type of code or structural coverage, SegTCP divides web pages into multiple segments representing distinct functional areas and treats these segments as key coverage targets. The method considers multiple coverage-related objectives simultaneously: maximizing coverage across different functional segments, minimizing redundant testing of sibling elements with similar functionality, diversifying the types of covered objects, and maximizing overall object coverage within each page. As a result, SegTCP achieves more balanced and comprehensive test case prioritization, effectively reducing redundancy while improving fault detection performance. Moreover, to address the multi-objective nature of the problem, SegTCP employs evolutionary search algorithms such as AGE-MOEA and NSGA-II to explore optimal test case orderings, thereby balancing trade-offs among competing objectives and identifying well-distributed Pareto-optimal solutions.

Rahman et al. \cite{Rahman2018PrioritizingDT} proposed a history-based test case prioritization method that uses dissimilarity analysis to improve regression testing effectiveness. Their approach combines historical fault data with inter-test-case dissimilarity to prioritize test cases that can detect faults early. Unlike traditional history-based methods that focus on similarity and often result in redundant fault detection, Rahman et al. introduced dissimilarity clustering. They first generate a Call Dependency Graph (CDG) to identify similar test case clusters and then construct a fault history CDG to rank test cases within each cluster based on connectivity and historical fault detection performance. The highest-priority test case from each cluster is selected to form a set of dissimilar clusters, which are then ordered to produce the final prioritized test suite. This strategy enhances fault detection by balancing historical data utilization with diversity through dissimilarity. REMAP \cite{PRADHAN201986} is a dynamic test case prioritization method that mines historical execution data to identify potential execution relationships among test cases. It integrates multi-objective optimization techniques to prioritize test cases statically and adjust their execution order dynamically at runtime, aiming to detect software faults earlier and improve regression testing efficiency. However, its performance is limited when historical data is insufficient. Parallel test case execution may also reduce its efficiency, and the accuracy of rule mining is sensitive to the confidence threshold setting, which needs to be adjusted for specific domains and data, thus lacking generalizability. Garg et al. \cite{Garg2024TestCP} proposed a test case prioritization method based on an improved NSGA-2 that leverages historical data to assess the sensitivity index (SI) of test cases. By considering fault detection capability, code coverage, and correlation with fault-sensitive regions, the method aims to maximize APFD and SI while minimizing execution cost, thereby improving regression testing efficiency and quality. However, it faces limitations in computational efficiency when handling large test case suites, especially when multiple complexity metrics are considered, leading to increased running time and reduced application efficiency in industrial settings.


\paragraph{\textbf{Test Suite Minimization (TSM)}}
Also known as test reduction, TSM \cite{harrold1993methodology} focuses on removing redundant or less impactful test cases while preserving overall fault detection ability or functional coverage. Minimization techniques often rely on greedy heuristics \cite{Sheikh2023AnOT, Cruciani2019ScalableAF,Zayed2021OptimizingTS, Bharathi2022OptimumTS} or clustering methods \cite{donaldson2021test, chaleshtari2024aim, viggiato2022identifying, alsharif2020sticcer,Kumar2013SoftwareTO} to identify and eliminate overlap among test cases.


TestReduce \cite{Sheikh2023AnOT} is a heuristic-based test case reduction method designed to optimize regression testing. This method integrates Genetic Algorithm (GA), a heuristic search technique, to construct an objective function that evaluates the importance of test cases by considering multiple dimensions, including requirement priority, requirement correlation, the extent of error correction in modules, and the correlation of corrected modules. Leveraging the global search and optimization capabilities of Genetic Algorithm, TestReduce efficiently screens out a minimized set of test cases from a large pool, thereby significantly reducing the testing workload and cost while ensuring test coverage and error detection capability.
FAST-R \cite{Cruciani2019ScalableAF} is a heuristic test case reduction method that efficiently selects representative subsets from large-scale test suites to reduce regression testing costs. It uses heuristic algorithms like k-means++ and importance sampling to identify uniformly distributed points in the test case space, ensuring key software behavior coverage without redundancy. Relying on test case textual information, FAST-R maps cases to a high-dimensional space and applies random projection for dimensionality reduction, enhancing computational efficiency. It is scalable and performs well in practice, especially with over 500,000 test cases, maintaining comparable fault detection capabilities while reducing test cases quickly.
Zayed et al. \cite{Zayed2021OptimizingTS} proposed an test case reduction method by integrating genetic algorithms with greedy algorithms. The genetic algorithm initially optimizes the test set to ensure coverage, while the greedy algorithm subsequently performs precise selection, leveraging its local optimality to identify the smallest test case subset. This method significantly reduces the number of test cases while maintaining high coverage, making it suitable for resource-sensitive software projects.
FTCBACO \cite{Bharathi2022OptimumTS} emulates ant foraging behavior to optimize test case selection. It maps test cases to nodes, using pheromone intensity as edge weights to identify the best test case combination for fault coverage while minimizing execution time. Initially, each test case is assigned an “ant” that selects paths based on pheromone levels and heuristic rules, with dynamic pheromone updates reinforcing high-quality paths. The test cases traversed by the fastest “ant” are selected as the optimal solution, reducing the test suite efficiently. However, the algorithm is sensitive to parameters like pheromone evaporation rate and heuristic factor, making parameter tuning crucial and increasing application complexity.

Kumar et al. \cite{Kumar2013SoftwareTO} proposed a fuzzy clustering-based test case reduction method to optimize test case selection while maintaining coverage and fault detection. Using the Fuzzy C-Means (FCM) algorithm, it groups similar test cases into clusters based on membership values, eliminating redundancy. The initial number of clusters is set according to program cyclomatic complexity, and then optimized by minimizing the standard deviation of cluster centers. A representative test case from each cluster forms the reduced suite, reducing test case count while maintaining good coverage. This method reduces testing workload and cost, enhances test case quality and efficiency through fuzzy clustering, and offers an innovative optimization strategy for software testing.
Pang et al. \cite{Pang2017ACT} proposed a clustering-based test case reduction technique aimed at reducing the cost of regression testing by categorizing test cases. The method utilizes coverage information from earlier versions of the program and combines k-means and hierarchical clustering algorithms to classify test cases into effective and ineffective categories, thereby avoiding the re-execution of non-effective test cases. Experimental results indicate that the method performs well in terms of recall, effectively identifying test cases that can expose faults. However, improvements are still needed in precision and accuracy. Moreover, the method's performance deteriorates when dealing with multiple faults, and for some programs, failing test cases are always assigned to different clusters, making binary clustering challenging.
Carmen et al. \cite{Coviello2018ClusteringSF} proposed a clustering-based approach for reducing test suites in software regression testing. The method groups test cases into clusters based on similarities in code coverage profiles, using metrics like cosine or Jaccard similarity. Within each cluster, the test case covering the most statements is selected as the representative for the reduced test suite. Experiments show that this approach significantly reduces test suite sizes while maintaining high fault detection effectiveness, with minimal degradation in some scenarios.

\paragraph{\textbf{Test Case Selection (TCS)}}
TCS \cite{rothermel1996analyzing} aims to select a relevant subset of test cases based on system modifications, enabling efficient retesting during incremental software updates. Static and dynamic program analysis methods are commonly employed to determine change-impacted regions and corresponding tests \cite{zhang2024hybrid, aghababaeyan2024deepgd, mehta2021data}.

Li et al.  \cite{Li2019MethodLevelTS} proposing a method-level test case selection technique named MEST. This approach aims to tackle the limitations of existing static methods within continuous integration environments. Traditional approaches predominantly depend on static analysis for identifying code dependencies; however, they frequently encounter challenges such as overlooked dependencies (e.g., dynamic calls in reflection) and imprecise dependency mapping (e.g., dynamic binding in inheritance). These issues can result in suboptimal test case selection, potentially omitting essential tests or including unnecessary ones. MEST integrates dynamic execution rules with static analysis, thereby enhancing the accuracy of method-level dependency identification. By examining dynamic behaviors associated with reflection and inheritance, MEST more precisely captures the required dependencies. Consequently, this not only boosts the precision of test selection but also significantly reduces the size of the test suite, while still maintaining high fault detection efficiency.
Fosse et al. \cite{Fosse2019SourceCodeLR} proposed an integrated model-driven regression test selection method combining static analysis and dynamic execution information. The method constructs an Impact Analysis Model by integrating static source code models with dynamic execution traces. Unlike traditional static methods, it incorporates dynamic execution trace information to reflect actual runtime behavior. The static model is generated using the MoDisco framework, and dynamic execution traces are injected into it to form the Impact Analysis Model. During regression testing, the method identifies change points and queries the model to determine which test cases executed these points, enhancing selection precision and reducing unnecessary test execution. The model is built offline, minimizing overhead and shortening regression testing time. Additionally, its reusability supports other tasks like fault localization and energy consumption analysis, enhancing practicality and extensibility.
DIRTS \cite{Hundsdorfer2023DIRTSDI} is a static test case selection tool that analyzes source code annotations and metadata to identify dynamic dependencies introduced by Dependency Injection (DI) frameworks such as Spring, Guice, and CDI, and incorporates them into a dependency graph. It extends the traditional dependency graph by adding DI-related edges to precisely track the impact of DI code changes on test cases. DIRTS supports both class-level and method-level test selection and can serve as an extension to existing regression test selection tools. However, its drawbacks include the high overhead of method-level analysis, which leads to extended execution time and may overestimate dependencies.

Sinaga et al. \cite{MarulituaSinaga_2019} proposed a dynamic test case selection method called Dynamic Partitioning with Additional Branch Coverage (DP-ABC). This method enhances software testing by dynamically adjusting test case selection based on feedback. It first partitions the test cases using an additional branch coverage algorithm to ensure diverse branch path coverage. Then, it dynamically categorizes test cases into ``good", ``medium", and ``poor" based on their past fault detection performance. Test cases are selected according to their category priority, with their categories being upgraded or downgraded based on current execution results. This dynamic adjustment optimizes test case selection, prioritizing those more likely to detect faults, thereby improving testing efficiency.
Wang et al. \cite{wang2023test} proposed uRTS (Unified Regression Test Selection), the first technique for Unified Regression Testing (URT) that considers both code and configuration changes. Unlike previous methods that handle these changes separately, uRTS integrates dependency analysis across both dimensions, selecting tests based on dynamic analysis of code and configuration dependencies. Using a two-dimensional selection mechanism, vertical (across code revisions) and horizontal (across configurations), uRTS reduces redundant test executions while maintaining safety guarantees. Empirical results on five large-scale open-source projects showed that uRTS reduced regression testing time by 3.64X compared to full testing and 1.87X compared to existing RTS methods.
Spaendonck \cite{Spaendonck2023EfficientDM} introduced a test case selection technique grounded in dynamic analysis, which leverages locally optimal choices to approximate globally optimal paths and circumvents revisiting states that have already been tested. Empirical evidence has shown that this method yields significant results in large and complex models. However, a limitation of this approach is that it may struggle to access all states in non-fully connected models, thereby potentially compromising test completeness due to suboptimal path selection.

While these studies demonstrate the maturity and diversity of test optimization research in traditional software systems, they largely operate under assumptions of static system architectures, deterministic input-output behavior, and homogeneous testing environments. These assumptions limit the direct applicability of traditional optimization techniques to ROS-based autonomous systems, where dynamic environmental interactions, asynchronous communications, and real-time constraints introduce fundamentally new complexities.

\subsection{Challenges in Optimizing Regression Testing for ROS-based Autonomous Systems}

In Section~\ref{tech_traditional}, we conducted an in-depth investigation into the various approaches and research findings of traditional software regression testing optimization techniques. These techniques have achieved remarkable progress in enhancing testing efficiency, reducing testing costs, and improving fault detection capabilities. For example, TCP enables earlier fault detection by optimizing the execution order of test cases; TCS reduces unnecessary test executions by accurately identifying test cases affected by changes; and TSM further optimizes the utilization of testing resources by eliminating redundant test cases. The successful application of these techniques has provided a robust foundation for regression testing optimization in traditional software systems.

However, despite their excellent performance in many traditional software systems, most regression testing optimization techniques were developed under the assumptions of static, deterministic input-output behavior and homogeneous testing environments. While these assumptions are reasonable for many traditional software systems, they no longer hold in the emerging field of ROSAS. ROSAS are characterized by multi-modal inputs, distributed execution, dynamic asynchronous behavior, real-time and safety constraints, and complex system architectures. These characteristics make it difficult to directly apply traditional regression testing optimization techniques.


\paragraph{\textbf{Multi-modal Inputs and Distributed Execution}}
Unlike conventional applications with well-defined, deterministic input-output mappings, ROSAS must process complex, heterogeneous sensor data (e.g., images, LiDAR scans, proprioceptive signals) and coordinate distributed node behaviors, complicating coverage measurement and fault detection. The multi-modal input in ROS is implemented through topics and services, involving various message types (such as sensor data, geometric poses) and communication patterns (asynchronous publish-subscribe, synchronous request-response) \cite{electronics13091762}. For example, sensor data (LiDAR scans, images) and control system commands may be transmitted through different topics, and the message types include nested structures and arrays. Data from different modalities exhibit significant differences in terms of timestamps, frequency, and spatial resolution. For instance, visual data is typically a continuous video stream, whereas audio data consists of discrete audio segments. How to precisely align these data to achieve synchronized processing is a major challenge. Moreover, multi-modal data originates from various sensors, each with distinct physical characteristics and data formats. Therefore, integrating these data into a unified semantic space so that the system can perform comprehensive understanding is an urgent issue to be resolved \cite{Baltruaitis2017MultimodalML, Gu2023EndtoEndMS, Pascher2023AdaptiXA, Cao2023MoPAMP}. ROSAS are typically composed of multiple heterogeneous nodes or modules, which vary in functionality and processing capabilities and support distributed execution. To operate efficiently, the system needs to coordinate communication between nodes and task allocation, which makes the overall architecture more complex. This is especially true when handling multi-modal inputs, where different types of data and their processing workflows need to be reasonably allocated to corresponding nodes to enhance system performance and scalability \cite{Toffetti2023ROSbasedRA}. Meanwhile, the data diversity and processing complexity brought by multi-modal inputs also pose higher challenges and requirements for the system's fault tolerance and operational reliability.

\paragraph{\textbf{Dynamic and Asynchronous System Behaviors}}
ROS-base autonomous systems exhibit dynamic environmental interactions and asynchronous communication patterns (e.g., topic publishing/subscribing), introducing substantial non-determinism that traditional static coverage-based or dependency-based methods are ill-equipped to handle \cite{Lauer2018ResilientCO}. For instance, in navigation tasks, the different consumption orders of multiple nodes for the same sensor data may trigger race conditions. This nondeterminism makes error reproduction and root cause analysis challenging, necessitating the introduction of additional synchronization monitoring nodes. The environment perception (e.g., SLAM) and decision logic (e.g., path planning) of autonomous systems are highly dependent on external inputs, and test cases are difficult to cover all possible scenario combinations \cite{Zhang2024AutomatedRT}. For example, physical environment uncertainties such as changes in lighting and movement of obstacles may affect sensor data, leading to inconsistent test results.

\paragraph{\textbf{Real-time and Safety-critical Constraints}}
Testing in ROSAS must account for strict timing and synchronization requirements, where even minor delays can cause system instability. Traditional test optimization techniques often neglect real-time guarantees, making them insufficient for safety-critical autonomous system validation. In ROSAS, regression testing needs to be able to monitor and verify the system's operating status in real-time. However, existing testing methods may not provide sufficient real-time feedback, especially in complex autonomous systems. Runtime verification is a method of analyzing and checking dynamic behavior during system operation, which can ensure that properties expressed in formal languages are not violated during runtime \cite{Caldas2024RuntimeVA}. However, this method may require adding monitors to the system, which can increase runtime overhead. Moreover, the design and implementation of the monitors themselves also need to be considered to ensure that they do not have a negative impact on system performance.

\paragraph{\textbf{Complex System Architecture with Stochastic Machine Learning Models and Rule-based Modules}}
ROS-based autonomous systems typically consist of multiple functional modules, including perception, prediction, planning, and control \cite{Zheng2025NeuroStrataHN}. Among them, perception and prediction modules heavily rely on stochastic machine learning models, which introduce inherent variability and non-determinism in system behavior due to probabilistic inference and data-driven decision-making. In contrast, planning and control modules are often governed by deterministic rule-based programs, which follow explicitly defined logic to ensure predictable and safe actuation responses. This heterogeneous architecture presents a significant challenge for traditional code coverage metrics, which focus on syntactic elements (e.g., statements and branches) but fail to capture the semantic coverage or the diverse behavioral patterns exhibited across real-world scenarios. As a result, current coverage-based optimization techniques are inadequate for fully assessing or optimizing the test quality in such systems.


These challenges necessitate the adaptation or reinvention of testing optimization strategies tailored to the ROSAS environment in order to meet the stringent demands of autonomous systems in terms of testing efficiency, fault detection capability, and resource utilization. Addressing these unique challenges forms the foundation for the techniques and insights discussed in the subsequent sections.


\section{Emerging Regression Testing Techniques on ROS-based Autonomous Systems} \label{sec:ros}

In Chapter~\ref{sec:traditional}, we provide a detailed review of traditional software regression testing optimization techniques, which have achieved significant progress in conventional software systems. However, with the rapid development of autonomous systems, particularly those based on ROS, these traditional techniques are confronted with new challenges. To address these challenges, numerous regression testing techniques have emerged in recent years. Although not specifically designed for ROSAS, these emerging techniques have demonstrated considerable potential in dealing with dynamic behaviors, multi-modal data, real-time constraints, and safety-critical aspects, which are precisely the unique challenges faced by ROSAS.

To address Research Question 1—``What are the major regression testing optimization techniques applicable to ROS-based autonomous systems?"—this chapter systematically reviews and analyzes traditional and emerging test optimization approaches. Given the unique characteristics of ROSAS, including dynamic behaviors, multi-modal sensor fusion, asynchronous operations, and safety-critical constraints, it is essential to reexamine how classical testing techniques can be adapted and extended.

\begin{figure*}[!ht]
	\setlength{\belowcaptionskip}{-0.3cm}
	\centering
	\includegraphics[width=\textwidth]{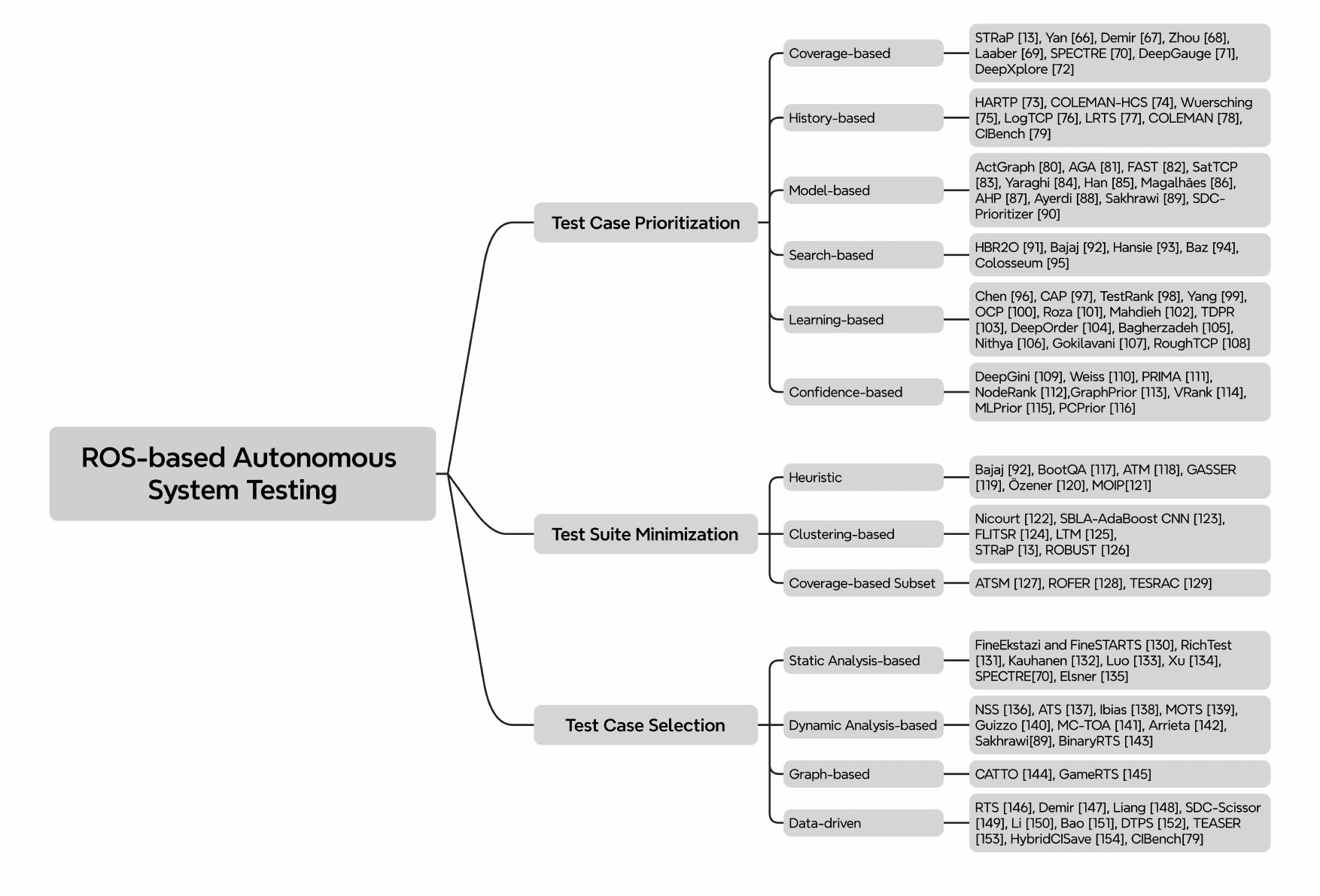}
	\caption{The research taxonomy of this survey.}
	\label{Taxonomy} 
\end{figure*}

Figure~\ref{Taxonomy} presents a taxonomy of the major test optimization techniques discussed in this survey. The taxonomy organizes methods into three primary categories: Test Case Prioritization, Test Suite Minimization, and Test Case Selection. Each category is further refined based on specialized strategies tailored to address the challenges specific to ROS-based autonomous systems, such as real-time performance, semantic diversity, and behavior-driven validation.

Based on this taxonomy, the following sections systematically examine each major category of optimization techniques. We first review TCP methods and their adaptations for autonomous systems, followed by discussions on TSM techniques, and TCS strategies that combine multiple optimization objectives.

\subsection{Test Case Prioritization}
Test case prioritization traditionally aims to reorder test cases to achieve specific objectives, such as maximizing early fault detection or improving coverage efficiency (as shown in Section \ref{tech_traditional}). In conventional software testing, TCP strategies often rely on static code analysis or fault history to determine execution order. However, when applied to ROS-based autonomous systems, TCP requires significant adaptation.

ROSAS present unique challenges for prioritization due to their highly dynamic and asynchronous nature. These systems must integrate heterogeneous sensor inputs (such as images, LiDAR scans, and proprioceptive data) across distributed computational nodes communicating via topics and services. Consequently, faults may emerge not merely from code errors, but from complex interactions between perception, planning, and control modules operating under real-time constraints. Prioritizing test cases in this context involves accounting for dynamic environment changes, node interdependencies, and multi-modal sensor fusion errors, aspects rarely considered in traditional TCP approaches.

Moreover, the temporal behavior of ROSAS where test outcomes may vary depending on timing, synchronization precision, and environmental conditions necessitates new prioritization criteria beyond static coverage metrics. As such, TCP in ROSAS must incorporate semantic awareness of the system’s operational context, real-time performance factors, and multi-source data interactions to remain effective.

\subsubsection{Categories of TCP Techniques}
Various strategies have been developed to prioritize test cases, aiming to optimize fault detection efficiency or maximize critical system behavior coverage. In the context of ROSAS, traditional TCP categories must be reinterpreted to account for the multi-modal, distributed, and real-time characteristics of these systems. The major categories of TCP techniques applicable to ROSAS testing include coverage-based, history-based, model-based, search-based, learning-based, and confidence-based prioritization.

\paragraph{\textbf{Coverage-based Prioritization}}
Coverage-based prioritization ranks test cases according to the extent to which they cover specific system elements. In traditional software testing, this often refers to code elements such as statements, branches, or paths \cite{yan2022test, demir2022dominating, zhou2021parallel, laaber2021applying}.

However, for ROS-based autonomous systems, coverage extends beyond static code to include: (1) dynamic coverage: runtime code execution paths \cite{laaber2021applying}; (2) sensor input variation: the range of environmental stimuli (e.g., different lighting, terrain, obstacle configurations) encountered \cite{lu2021search}; and (3) neuron coverage: the activation status of individual neurons, the activity level of multi-layer neurons, and the coverage of boundaries and thresholds \cite{Ma2018DeepGaugeMT, Pei2017DeepXploreAW}.

TCP based on coverage continues to evolve in multiple fields.  Zhou et al. \cite{zhou2021parallel} proposed a parallel prioritization framework that combines code coverage with multi-threaded acceleration, significantly reducing the testing time for large-scale systems. Demir et al. \cite{demir2022dominating} transformed the problem of test case prioritization into finding the minimum dominating set by constructing a directed bipartite graph from the set of test cases to the set of requirements. The algorithm sorts test cases based on the number of requirements they cover, prioritizing those that cover the most uncovered requirements, thereby achieving efficient coverage of all requirements. This method is particularly suitable for requirement-driven testing scenarios, maximizing software requirement coverage within limited time and resources, and improving the efficiency and effectiveness of regression testing.

A representative study STRaP \cite{deng2022scenario} introduces a scenario-based vectorization method that encodes semantic features such as dynamic entities (vehicles, pedestrians, behaviors) and static entities (traffic lights, stop signs) into integer vectors. STRaP segments driving records into semantic fragments by comparing consecutive vector similarities and employs a smoothing mechanism via sliding windows. It prioritizes tests based on coverage and rarity, assigning higher weights to infrequent features to detect faults more effectively. This shows that prioritizing test cases that trigger a wide range of node interactions and sensor input conditions can lead to more effective fault discovery in ROSAS.

DeepGauge \cite{Ma2018DeepGaugeMT} quantifies the activation patterns of test inputs on deep neural networks from multiple perspectives, such as the range of neuron outputs, boundary values, and extreme activation states. It measures the degree to which test data covers the internal behavior of deep learning models, thereby comprehensively evaluating the adequacy of testing. DeepXplore \cite{Pei2017DeepXploreAW} guides the generation and prioritization of test cases by measuring coverage metrics such as neuron activations, improving test coverage and fault detection capabilities. Yan et al. \cite{yan2022test} proposed a class-specific pattern extraction method based on the distribution of pre-activation output values as a new coverage criterion to evaluate the behavioral characteristics of deep neural networks. By comparing the degree of difference between test cases and these patterns, test cases are prioritized, with potential error-triggering cases that deviate from expected patterns being handled first. This method not only focuses on whether neurons are activated but also utilizes the distribution of pre-activation valuations, providing a more detailed coverage perspective to improve the effectiveness and efficiency of model testing. Lu et al. \cite{lu2021search} prioritize test scenarios by quantifying their coverage of critical system functions, selecting those that cover more important functions first.

\paragraph{\textbf{History-based Prioritization}}
History-based prioritization uses past execution data such as fault detection rates or test effectiveness to inform test ordering \cite{wang2024hierarchy, prado2022cost}. In ROSAS, this approach can leverage: (1) historical failure records associated with specific nodes or modules (e.g., localization, perception failures) \cite{wang2024hierarchy}; (2) frequency and severity of prior faults under certain environmental conditions \cite{wuersching2023severity}; and (3) system logs indicating instability patterns correlated with specific message flows or synchronization events \cite{chen2023exploring}. Integrating such historical insights allows testers to prioritize scenarios more likely to reveal critical faults in complex autonomous operations.

HARTP \cite{wang2024hierarchy} utilizes historical data, including test execution information, fault detection records, and code coverage, to calculate priority scores for each test class, guiding regression test prioritization. These scores reflect the “importance” of tests in fault detection and, combined with a hierarchy-aware (HA) method, sort the test modules and test cases within modules to reduce testing runtime costs. Cheng et al. \cite{cheng2024revisiting} revisited the problem of test case prioritization in long-running test suites, evaluating and improving the effectiveness of various history-based prioritization algorithms in this context. Their results indicated that test case prioritization strategies relying on test execution time and recent failure records perform optimally in long-running test suites.

Wuersching et al. \cite{wuersching2023severity} proposed a severity-aware prioritization approach focusing on system-level regression testing in automotive software. This method comprehensively considers the historical fault detection performance of test cases and the severity levels of faults, aiming to more accurately identify potential critical issues. COLEMAN \cite{lima2020multi} leverages historical data collected from each test execution in the CI environment to compute reward values for each test case. It evaluates the importance and fault detection capability of test cases through reward functions (such as RNFail based on failure frequency or TimeRank based on time ranking). Then, based on a multi-armed bandit (MAB) strategy, it dynamically selects the prioritization order of test cases, balancing the trade-off between exploring new test cases and exploiting known high-value test cases. After each test execution, the system updates the historical data based on the latest results, forming a feedback loop to continuously optimize test case prioritization. Furthermore, Prado Lima et al. \cite{prado2022cost} evaluated the fault detection capability of test cases through a credit assignment mechanism based on historical failure data. They used a sliding window to record test results from the most recent builds, combined with a multi-armed bandit (MAB) framework, to balance exploration of new test cases and exploitation of known efficient test cases based on historical data. This method can adaptively optimize test order in continuous integration environments solely based on historical data, without requiring additional information such as code coverage.

LogTCP \cite{chen2023exploring} initially preprocesses and parses historical execution logs to extract events and patterns related to test cases. Based on the features generated from these log data (such as event frequency, event sequences, or event relationship graphs), it performs prioritization evaluation and sorting of test cases. Jin and Servant \cite{jin2021cibench} introduced CIBench, a dataset and benchmarking framework for systematically evaluating test and build selection and prioritization techniques in CI environments. By replicating and simulating 14 variants of 10 techniques across multiple projects, CIBench offers insights into optimizing both time-to-feedback and computational cost through fine-grained test prioritization and selection. Although designed for CI workflows, its approach (particularly the integration of dependency analysis, build logs, and historical failure patterns to guide prioritization) provides valuable implications for autonomous systems testing. In such systems, where frequent builds, dynamic software updates, and resource constraints are also prevalent, CIBench-inspired techniques can help prioritize tests that are most likely to expose critical failures while reducing testing overhead. This makes it particularly relevant to advancing test case prioritization strategies for ROSAS operating under similar continuous deployment and real-time validation requirements.

\paragraph{\textbf{Model-based Prioritization}}
Model-based prioritization relies on abstract representations of system behavior, such as finite state machines, behavior trees, or task graphs \cite{chen2023actgraph, li2021aga, chen2024fast, li2024semantic, yaraghi2022scalable, han2020convergence, magalhaes2021ui, nayak2022analytic, ayerdi2022multi}. In the context of ROSAS, model-based TCP can involve: (1) model activation graph-based approach: constructing an activation graph to extract node features and establish a prioritization model, which enables the prioritization of test cases; (2) rule mining and multi-objective optimization-based method: leveraging rule mining and search algorithms in conjunction with multi-objective optimization techniques to dynamically prioritize test cases based on their execution outcomes. Model-based approaches are particularly valuable in ROSAS due to the high-level decision-making and mission-critical tasks that must be validated across diverse operational contexts.


ActGraph \cite{chen2023actgraph} records neuron activation values and constructs an activation graph, then extracts higher-order node features to characterize the activation patterns of test cases. Finally, it prioritizes test cases based on these features, selecting those with features similar to known error behavior patterns first, thereby more effectively detecting potential errors in the model. FAST \cite{chen2024fast} combines feature selection and uncertainty estimation to optimize test case prioritization. Initially, it quantifies the importance of each feature to model predictions through feature selection algorithms and dynamically prunes noisy features during inference to enhance the accuracy of uncertainty estimation. Subsequently, it derives new probability vectors using the pruned features and prioritizes test cases based on their uncertainty levels, with higher uncertainty cases executed first to more effectively detect potential errors. SatTCP \cite{li2024semantic} is a semantics-aware two-stage method. In the coarse-grained filtering stage, it employs natural language processing techniques to calculate the semantic similarity between code changes and test cases, rapidly screening high-priority test cases. In the fine-grained ranking stage, it combines static analysis and dynamic execution data to conduct multi-dimensional feature evaluation and precise ranking of the filtered test cases, thereby significantly improving defect detection rates and reducing execution time costs. Magalhães et al. \cite{magalhaes2021ui} conducted research on the prioritization of UI test cases in an industrial environment, providing practical prioritization criteria and methods for domain-specific testing.  Sakhrawi et al. \cite{sakhrawi2024test} employed a model-based TCP on Ontology and COSMIC measurement. Initially, they constructed an Ontology model for semantic reasoning to identify test cases related to software changes. Subsequently, they quantified the functional size of these test cases using the COSMIC method and calculated their priorities in combination with risk and historical fault data.

Birchler et al. \cite{birchler2023single} proposed SDC-Prioritizer, a simulation-based TCP framework specifically targeting self-driving cars, which are representative autonomous systems with highly dynamic and safety-critical behaviors. Recognizing that traditional blackbox prioritization techniques are unsuitable for such complex virtual environments, the authors designed SDC-Prioritizer to prioritize virtual driving tests using static features derived from road networks, which can be computed prior to test execution. Two variants were introduced: SO-SDC-Prioritizer, a single-objective genetic algorithm prioritizing based on diversity and cost, and MO-SDC-Prioritizer, a multi-objective approach using NSGA-II to balance fault-revealing potential and execution efficiency. Empirical results on multiple datasets demonstrated that MO-SDC-Prioritizer significantly improves early fault detection with minimal computational overhead. Importantly, its reliance on static, pre-execution features makes this approach highly suitable for ROSAS, where real-world and in-field testing are costly and risky, and where multi-modal sensor-driven scenarios require efficient prioritization under limited computational and time budgets.

\paragraph{\textbf{Search-based Prioritization}}
Search-based prioritization typically transforms the problem into a multi-objective optimization problem, employing metaheuristic algorithms (such as genetic algorithms and particle swarm optimization) to seek approximate optimal solutions. In the context of ROSAS, search-based prioritization utilizing heuristic search algorithms (such as greedy algorithms, metaheuristics, and evolutionary algorithms) to optimize the prioritization of test cases, thereby enhancing efficiency and effectiveness \cite{raamesh2022cost, bajaj2022improved, mondal2021hansie, baz2024prioritizing,mondal2021sf}. Given that ROSAS operates in dynamic environments (such as obstacles and sensor noise) and often faces multi-objective conflicts (such as real-time performance, resource utilization, and task completion), search-based methods hold significant application value in ROSAS.




HBR2O \cite{raamesh2022cost} is a hybrid battle royale-based remora optimization search algorithm that integrates the adhesion-following mechanism of the remora optimization algorithm and tactical competitive strategies. By simulating symbiotic relationships and competitive-cooperative processes, it optimizes the selection and prioritization of test cases. HBR2O first employs the adhesion mechanism to select test cases with strong complementarity to enhance coverage. Subsequently, it dynamically adjusts priorities through tactical competitive mechanisms to achieve a globally optimal combination of test cases. Similarly, the improved bat algorithm by Bajaj et al. \cite{bajaj2022improved}, which leverages the simulation of animal characteristics, optimizes the order of test cases, thereby improving test coverage and fault detection rates. Hansie \cite{mondal2021hansie} combines multiple strategies, employing a hybrid and consensus approach for TCP. This algorithm integrates various prioritization strategies, such as those based on code coverage, historical execution outcomes, and change impact analysis. By consolidating the strengths of different strategies, it enhances the comprehensiveness and robustness of the prioritization. Hansie also introduces ranking aggregation techniques from social choice theory to reach a consensus on the ordering of multiple prioritization results, thereby generating the final sequence for test case execution to more rapidly detect defects and improve testing efficiency. Furthermore, Baz et al. \cite{baz2024prioritizing} proposed a search algorithm based on test dependencies and execution time to optimize test ordering. This model analyzes the dependencies between test cases to ensure that dependent test cases are executed first, while also evaluating the execution time of each test case to prioritize those with shorter execution times. Colosseum \cite{mondal2021sf} is a heuristic search algorithm based on Delta displacement. Its core involves calculating three displacement parameters of code changes (Delta) in the execution path of test cases (initial offset, average dispersion, and terminal offset) to generate a “total displacement value.” Test cases are prioritized based on this value, with smaller displacement values indicating a higher likelihood of exposing defects and thus prioritizing their execution.

\paragraph{\textbf{Learning-based Prioritization}}
In recent years, learning-based prioritization has emerged as a promising direction for enhancing TCP by leveraging machine learning and deep learning techniques to predict the fault detection potential or importance of test cases. Unlike traditional heuristics or search-based approaches, which rely heavily on manually defined criteria such as code coverage or modification impact, learning-based methods can automatically learn prioritization patterns from historical testing data, execution traces, or code changes \cite{ chen2022focus, wang2024cluster, li2021testrank, yang2023can, zhang2022test, da2022machine, mahdieh2022test, shen2024prioritizing, sharif2021deeporder, bagherzadeh2021reinforcement, nithya2023fuzzy, gokilavani2021test}.

In the context of ROSAS, learning-based prioritization holds particular promise. Autonomous systems typically generate rich and diverse data during operation and testing, including logs, sensor streams, and multi-modal outputs. Learning-based approaches can be trained on this data to predict which test scenarios are most likely to trigger safety-critical failures or exercise rarely tested behaviors \cite{li2021testrank}. Learning-based methods are generally classified into three categories: (1) unsupervised learning, which aims to discover the latent structure and distribution characteristics within data \cite{gokilavani2021test, mahdieh2022test}; (2) semi-supervised or supervised learning, which leverages labeled data to improve the accuracy of predictive models \cite{li2021testrank}; (3) reinforcement learning, which dynamically optimizes decision-making strategies through feedback signals obtained via interaction with the environment \cite{bagherzadeh2021reinforcement}. Regardless of the learning paradigm adopted, improvements in testing efficiency and defect detection capability can be achieved to varying extents. Moreover, as autonomous systems evolve frequently through continuous integration and deployment, learning-based prioritization techniques can be updated incrementally to reflect the latest system behavior patterns, providing adaptive and context-aware prioritization strategies.


Gokilavani et al. \cite{gokilavani2021test} proposed a TCP method based on principal component analysis (PCA) and K-means clustering. Initially, PCA was employed to reduce the dimensionality of test case features, extracting key characteristics. Subsequently, K-means clustering was utilized to categorize test cases into distinct clusters. Finally, inter-cluster and intra-cluster TCP were conducted based on critical features such as fault detection rate and coverage, thereby generating the ultimate prioritization sequence. Roza et al. \cite{da2022machine} employed a sliding window to filter test cases, followed by training a random forest model to predict the fault probability of test cases, and subsequently prioritized the test cases according to the predicted scores. When the time budget was set at 10\%, the model demonstrated satisfactory performance. However, under ample time budget conditions, its performance failed to reach the level of the previous baseline model. Mahdieh et al. \cite{mahdieh2022test} introduced a prioritization approach that integrates test case diversity and fault proneness estimation. This method initially utilized clustering analysis to assess the diversity of test cases, ensuring that test cases covering different code regions were prioritized for execution. Subsequently, a machine learning model was employed to estimate the fault proneness of test cases, predicting which ones were more likely to uncover defects. By combining these two dimensions (diversity and fault proneness), the final prioritization sequence was generated. The innovation of this method lies in its simultaneous consideration of both the breadth of test case coverage and their defect detection capability, thereby overcoming the limitations of traditional methods that focus solely on a single dimension.

Nithya et al. \cite{nithya2023fuzzy} proposed a hybrid model for TCP in regression testing, based on fuzzy logic and artificial neural networks (ANN). Initially, fuzzy logic rules were applied to fuzzify the attributes of test cases, such as historical fault rate, requirement coverage, and modification impact scope. The fuzzified features were then input into the ANN to predict the fault probability of test cases and prioritize them accordingly. The innovation of this method lies in its combination of the flexibility of fuzzy logic and the predictive power of ANN, enabling the handling of uncertainties and fuzziness in test case attributes. CAP \cite{wang2024cluster} is an adaptive TCP method based on clustering analysis. By integrating clustering algorithms and adaptive algorithms, it dynamically adjusts the priority of test cases during the test execution process, thereby enhancing fault detection efficiency. This method addresses the challenge of TCP lag in regression testing, which arises due to the inability to promptly utilize fault information during the test execution period. RoughTCP \cite{guaceanu2024leveraging} proposes an unsupervised TCP method using rough sets theory and agglomerative clustering for dynamic, data-intensive CI environments. Unlike supervised methods that need labeled data or neural models prone to overfitting, RoughTCP clusters and prioritizes test cases based on intrinsic attributes like duration and fault rate. It adapts without retraining and handles uncertain cases through rough set approximations. Evaluated on three industrial datasets, it outperforms state-of-the-art methods at medium and high testing budgets (75\% and above). This makes RoughTCP suitable for ROSAS environments, which lack labeled data and experience frequent updates. Its ability to dynamically group and reprioritize test cases ensures critical paths remain tested, aligning with scenario-based and semantic-aware testing strategies.

TestRank \cite{li2021testrank} employs a semi-supervised learning approach to prioritize unlabeled test instances. This method constructs a similarity graph between test instances based on graph neural networks (GNN) and subsequently extracts contextual features from a small amount of labeled data through semi-supervised learning. The core innovation of TestRank lies in its integration of the intrinsic features of test instances (such as model prediction confidence) with contextual features (such as similarity to other test instances) to predict their fault-revealing capability. DeepOrder \cite{sharif2021deeporder} is a TCP method for continuous integration testing based on deep learning. This method trains a regression model using historical test data to automatically learn the relationship between various attributes of test cases (such as execution time, historical execution status, and requirement coverage) and fault probability. It then predicts the fault probability of test cases and generates a prioritization sequence accordingly. OCP \cite{zhang2022test} is a TCP method based on partial attention mechanism. It focuses on the key features of test cases and employs a local attention mechanism to weight the features of test cases, highlighting important features. This enables a more accurate assessment of the fault detection potential of test cases and achieves prioritization. The innovation of this method lies in its introduction of the attention mechanism to focus on the key coverage parts of test cases, thereby improving the accuracy of prioritization. TDPR \cite{shen2024prioritizing} is a TCP method for deep neural networks (DNNs)-based on training dynamics. This method analyzes the dynamic features during the DNN training process, such as neuron activation patterns and gradient changes, to identify test inputs that have a significant impact on model training. It then prioritizes these test inputs accordingly. The authors designed an unsupervised learning framework that can automatically capture key patterns in training dynamics to predict the importance of test inputs. The innovation of this method lies in its use of training dynamics as a new perspective to evaluate the importance of test inputs, whereas traditional methods mainly rely on features such as coverage during testing.

Bagherzadeh et al. \cite{bagherzadeh2021reinforcement} proposed a test case ranking method based on reinforcement learning (RL), modeling TCP problem as an RL problem. They employed three different ranking models, pairwise, listwise, and pointwise, to guide the interaction of the RL agent with the CI environment. Experimental results indicated that the pairwise ranking model combined with the ACER algorithm (pairwise-ACER) performed the best on rich datasets, with its ranking accuracy approaching the optimal strategy. This method can dynamically adapt to changes in the system and test suite within the CI environment while maintaining high test case ranking precision. Chen et al. \cite{chen2022focus} introduced a method for focusing on new test cases in CI testing based on reinforcement learning. Considering the frequent iteration of historical information in CI, they employed sliding window techniques to capture effective information. They introduced two methods for calculating rewards: fixed-size sliding window and dynamic sliding window. Empirical studies were conducted on fourteen industrial-level programs, and the results showed that the reward function based on the sliding window could effectively improve the TCP outcome. Specifically, the method based on the dynamic sliding window ranked 74.18\% of the failed test cases within the top 50\% of the ranking sequence. The core innovation of this method lies in training a reinforcement learning model with historical test data, enabling it to predict which new test cases are more likely to detect undetected defects. Yang et al. \cite{yang2023can} explored the impact of code representation techniques on information retrieval (IR) for TCP. The authors argued that traditional IR methods primarily rely on text similarity to determine TCP but neglect the structural and semantic information of the code. Therefore, they utilized deep learning models (such as BERT) to transform test cases and requirement specifications into vector representations. Subsequently, they prioritized the test cases based on vector similarity, thereby optimizing the TCP outcome and enhancing fault detection capability.

\paragraph{\textbf{Confidence-based Prioritization}}
The confidence-based prioritization is an approach that involves calculating and evaluating the confidence of test cases, and then ordering these test cases based on their confidence values to optimize the sequence of test execution. Confidence typically reflects the reliability and effectiveness of a test case in detecting defects or verifying software functionality \cite{Feng2019DeepGiniPM,Weiss2022SimpleTW,wang2021prioritizing,li2024gnn,dang2023graphprior,li2024prioritizing,dang2024test}.

In the context of ROSAS, automated systems typically incorporate numerous deep learning modules, which contain a vast number of neurons, making traditional software engineering testing methods less effective. In recent years, confidence-based test case prioritization techniques have emerged as promising approaches for testing deep learning systems.

DeepGini \cite{Feng2019DeepGiniPM} evaluates the model's uncertainty in classifying test cases by calculating the Gini impurity of the test cases. A higher Gini impurity indicates a lower confidence of the model in classifying the test case, implying a higher likelihood of misclassification, thereby enabling the prioritization of test cases.The primary advantages of DeepGini lie in its simplicity and computational efficiency; however, its performance is limited in complex tasks, dynamic environments, and certain specific error patterns such as high-confidence errors. PRIMA \cite{wang2021prioritizing} is a test input prioritization method that integrates mutation analysis (MA) and Learning-to-Rank. Initially, PRIMA employs mutation operators to generate multiple mutation models and corresponding mutation inputs. It then evaluates the distinguishing capability of test inputs against these mutation models and the distinguishing capability of the models against the mutation inputs, termed as the “surprise value.” This surprise value reflects the degree of output inconsistency of test inputs between the original model and mutation models, as well as the degree of output inconsistency of the model between the original input and mutation inputs, thereby quantifying the potential value of the test input. Subsequently, PRIMA trains a ranking model based on these evaluation results to predict the likelihood of test inputs exposing defects. Finally, new test inputs are scored by the ranking model, and inputs with high scores are prioritized for testing. Moreover, PCPrior \cite{li2024test} constructs a comprehensive set of feature vectors by integrating statistical characteristics of test data, mutation analysis results, and input uncertainty information quantified through methods such as DeepGini, entropy, and PCS. This feature vector set is then utilized to train a ranking model for prioritizing test cases.

\subsubsection{Discussion and Remarks}
Test case prioritization plays a critical role in optimizing the testing process for ROS-based autonomous systems. While traditional TCP techniques such as coverage-based, history-based, and model-based prioritization provide a valuable foundation, direct application to ROSAS environments is insufficient due to the systems' dynamic, distributed, and real-time characteristics.

In ROSAS, prioritization must consider multi-modal sensor interactions, asynchronous node communication, non-deterministic behavior, and real-world operational semantics. Emerging techniques from cyber-physical systems and autonomous driving domains, such as semantic vectorization of scenarios and behavior-driven test prioritization, offer promising directions for addressing these challenges. However, significant research gaps remain, particularly in developing context-aware, environment-sensitive, and real-time-capable prioritization frameworks tailored to ROS architectures.

Future work should focus on hybrid TCP strategies that integrate both structural coverage and semantic coverage metrics, leverage system logs and environmental context, and dynamically adapt to evolving operational conditions. Incorporating neurosymbolic reasoning and foundation models into TCP may further enhance the prioritization of test cases in complex, uncertain, and safety-critical autonomous systems.

\subsection{Test Suite Minimization}

Test suite minimization aims to reduce the size of a test suite by eliminating redundant or less valuable test cases, while maintaining essential testing objectives such as code coverage, fault detection capability, or system behavior validation. As summarized in Section \ref{tech_traditional}, TSM techniques have been widely employed in traditional software engineering to lower testing costs, accelerate feedback cycles, and improve resource utilization, particularly in continuous integration and continuous deployment (CI/CD) environments.

In the context of ROSAS, the need for effective TSM becomes even more critical. Autonomous systems typically involve extensive testing across diverse environmental conditions, multi-modal sensor inputs, and distributed computational nodes. Executing large-scale test suites on real hardware or high-fidelity simulation platforms can be prohibitively time-consuming and resource-intensive.

However, applying conventional TSM strategies directly to ROSAS presents unique challenges. Redundancy in ROSAS testing is not limited to code paths but also involves sensor configurations, inter-node communication patterns, and dynamic environmental interactions. Simple syntactic or structural minimization techniques risk overlooking subtle but important behavioral variations arising from asynchronous operations or multi-source data fusion.

Therefore, TSM in ROSAS environments require to evolve beyond traditional structural coverage preservation. It must incorporate semantic and contextual factors such as environmental diversity, sensor modality interactions, and timing dependencies to ensure that minimized test suites remain representative, behaviorally rich, and effective for validating safety-critical autonomous functionalities.

\subsubsection{Categories of TSM Techniques}
Several categories of TSM techniques have been developed to systematically reduce the size of test suites while preserving critical testing objectives. In the context of ROSAS, these techniques require careful adaptation to account for the complexity of multi-modal sensor inputs, distributed node architectures, and real-world dynamic environments. This section reviews the primary categories of TSM techniques and discusses their relevance and limitations when applied to ROSAS.

\paragraph{\textbf{Greedy Minimization}}
Greedy minimization techniques iteratively select test cases that maximize coverage or fault detection criteria until all required testing objectives are satisfied. Typically, a coverage matrix is constructed, mapping test cases to coverage elements (e.g., statements, branches, sensor conditions), and the algorithm greedily selects the test case covering the most uncovered elements at each step \cite{wang2024test, pan2023atm, coviello2022gasser, bajaj2022improved, ozener2020effective}.


In ROSAS, greedy approaches can be extended beyond code coverage to include: (1) mathematical modeling and optimization: employing formal modeling and mathematical optimization techniques to address the problem of test case reduction \cite{ozener2020effective, xue2020multi}; and (2) evolutionary computation and search strategies: algorithms or intelligent search strategies that simulate the process of natural evolution to automatically explore high-quality subsets of test cases. However, greedy strategies often assume static relationships between test cases and coverage elements. In dynamic ROS environments, where environmental changes and asynchronous behaviors impact test outcomes, greedy minimization must be augmented with semantic and temporal awareness.


Özener et al. \cite{ozener2020effective} proposed a multi-criteria TSM method by transforming the test case selection problem into a multi-objective optimization problem, aiming to identify optimal solutions through mathematical techniques such as linear programming. The optimization model considers multiple factors including code coverage, execution time, and resource consumption and selects the most suitable subset of test cases for specific testing scenarios based on varying weight assignments. However, this approach relies on linearizing the originally complex, potentially non-linear optimization problem, which may result in the loss of important relationships among criteria. Furthermore, the method requires domain expertise for determining appropriate weight allocations and lacks an automated mechanism for weight assignment, thereby increasing its complexity and limiting its practical applicability. Xue and Li \cite{xue2020multi} proposed a set of multi-objective integer programming (MOIP) approaches to address the multi-criteria test suite minimization (MCTSM) problem, which aims to balance trade-offs among test suite size, statement coverage, and fault detection capabilities. Unlike traditional weighted-sum methods that only produce a single solution and often fail to reflect the complete Pareto front, the proposed MOIP-based formulations, are designed to produce sound (Pareto-optimal) and complete sets of solutions. The experiments on multiple real-world subject programs demonstrated that MOIP significantly outperformed existing heuristic and evolutionary algorithms in both solution quality and efficiency, especially in scenarios requiring comprehensive trade-off analysis. For ROSAS, which typically operate under tight resource constraints while needing to ensure broad behavioral coverage and high fault detection effectiveness, this method is particularly relevant. The ability of MOIP to explicitly model and optimize multiple objectives makes it highly suitable for selecting minimal yet diverse and fault-revealing test subsets, which is critical in multi-sensor, behavior-rich autonomous environments where testing costs and risks must be tightly controlled.

GASSER \cite{coviello2022gasser} is a test case set reduction method based on multi-objective evolutionary algorithms, employing the NSGA-II algorithm to simultaneously optimize multiple objective functions, including maximizing statement coverage, maximizing test case diversity, and minimizing the size of the test case set. The core innovation of GASSER lies in its multi-objective optimization framework, which can balance the conflicting goals of test case coverage, diversity, and quantity. The NSGA-II algorithm, through non-dominated sorting and crowding distance calculation, is able to identify a set of Pareto optimal solutions in the solution space, allowing testers to select the most appropriate test case set according to specific needs. However, the high computational complexity of the NSGA-II algorithm, which requires a large number of iterations to converge to the optimal solution, may become a bottleneck when dealing with extremely large test sets. ATM \cite{pan2023atm} is a blackbox TSM method based on code similarity. By analyzing the code similarity between test cases, ATM identifies and removes redundant test cases, while further optimizing the test set through evolutionary search. The core innovation of ATM is its code similarity measurement mechanism. The authors designed a similarity measurement method based on the structure and semantics of test code, which can more accurately identify functionally similar test cases. Additionally, ATM introduces an evolutionary search strategy, which optimizes the test case set through operations such as crossover, mutation, and selection, minimizing the number of test cases while maintaining coverage. However, it may fail to accurately identify redundancy for test cases that are functionally similar but implemented differently, and may be inefficient when dealing with extremely large test sets. BootQA \cite{wang2024test} is a test case reduction method based on quantum annealing, modeling the test case selection problem as a quadratic unconstrained binary optimization (QUBO) problem and leveraging the global optimization capability of the quantum annealing algorithm to find the optimal solution. The quantum annealing algorithm, an optimization method based on quantum mechanics principles, aims to escape local optima through the quantum tunneling effect. The authors transformed the optimization objective of test case selection into the energy function of a quantum system, where the lowest energy state corresponds to the optimal test case set. When dealing with large-scale test sets, the global search capability of quantum annealing enables it to find better solutions, whereas traditional algorithms often get trapped in local optima. However, the QUBO modeling of the test case selection problem requires specialized knowledge, increasing the implementation complexity.

\paragraph{\textbf{Clustering-based Minimization}}
Clustering-based minimization groups similar test cases into clusters based on predefined similarity metrics (e.g., input profiles, output behaviors, sensor activation patterns) and selects representative tests from each cluster. This approach is particularly useful when the test space exhibits significant redundancy due to variations in environmental parameters or sensor noise \cite{nicourt2024using, raamesh2022test, callaghan2023improving, pan2023ltm}.


In ROSAS, clustering can be applied to: (1) functional behavior: clustering by leveraging the similarity of test cases in program behavior (such as covered variants, killing power, etc.) \cite{nicourt2024using, callaghan2023improving}; and (2) semantic representation: mapping test code into embedded representations in high-dimensional vector space, and then clustering and filtering based on vector similarity \cite{Timperley2024ROBUST2B, pan2023ltm}. Clustering-based methods offer flexibility in handling complex, multi-dimensional input spaces typical of ROSAS. However, defining effective similarity metrics that capture both functional behavior and safety-critical distinctions remains a challenging research problem.

Nicourt et al. \cite{nicourt2024using} proposed a TSM method based on hierarchical clustering and mutation testing. Specifically, after the clustering process yields result clusters, the test case that covers the most mutants or has the smallest kill rate is selected from each cluster as the representative. This method aims to mitigate the negative impact of accidental correctness (i.e., the program has errors but passes the test) on test reliability. However, this method also faces some challenges, such as the high computational overhead of mutant generation and the reliance on a comprehensive unit testing base. FLITSR \cite{callaghan2023improving} constructs a basis that can explain test failures by iteratively selecting the “most suspicious elements” and removing their covered failing tests. Each basis provides an explanation for test failures, and developers can select the most appropriate fault localization result based on these different explanations.

Pan et al. \cite{pan2023ltm} proposed LTM (Language model-based Test suite Minimization), a scalable and blackbox minimization approach that is the first to leverage large language models (LLMs) for TSM. Unlike traditional whitebox or syntax-based similarity techniques, LTM generates test method embeddings using pre-trained language models such as CodeBERT, UniXcoder, and CodeLlama to capture semantic and contextual relationships among test cases. Combined with vector-based similarity measures (Cosine and Euclidean) and a genetic algorithm, LTM efficiently selects a reduced yet diverse subset of tests. Empirical evaluation on large-scale open-source projects demonstrated that LTM achieves superior fault detection rates and minimizes test suites up to five times faster than state-of-the-art methods, especially benefiting larger and more complex test suites. For ROSAS, which often operate in blackbox scenarios and generate diverse test cases from multi-modal and mission-critical operations, LTM’s ability to perform similarity-based minimization without requiring source code or instrumentation offers clear advantages. Its scalability, blackbox nature, and semantic-awareness make it particularly well-suited for selecting representative tests while managing the execution cost and maintaining fault detection effectiveness in dynamic and evolving autonomous system environments.

\paragraph{\textbf{Coverage-based Subset Selection}}
Coverage-based subset selection focuses on retaining a subset of test cases that collectively satisfy specified coverage criteria. Traditional metrics such as statement or branch coverage can be used \cite{alekseev2024atsm}. In ROSAS environments, richer coverage models are necessary, including: (1) semantic coverage: ensuring tests collectively cover different high-level behaviors (e.g., obstacle avoidance, path planning recovery); and (2) sensor input space coverage: capturing variability across different sensor modalities and environmental conditions \cite{Bai2024MultiDimensionalAM}. By moving beyond purely structural coverage to incorporate system behavior and environment interaction models, coverage-based subset selection can produce minimized test suites that remain effective in validating the functional correctness and robustness of ROSAS.


ATSM \cite{alekseev2024atsm} is a TSM method based on code coverage. By quantifying the coverage effect of test cases on the code (such as statement coverage and branch coverage) and combining it with a scalable minimization engine (supporting strategies like greedy algorithms and genetic algorithms), it systematically identifies and removes redundant test cases. This significantly reduces the size of the test suite while ensuring the capability of defect detection. The method emphasizes a coverage-driven decision-making mechanism, aiming to balance the consumption of testing resources and the need for quality assurance. It is applicable to the CI scenarios of large-scale industrial software projects.

Becho et al. \cite{becho2022tesrac} proposed TESRAC, a comprehensive framework for systematically assessing and comparing TSM tools at scale. Unlike prior studies focusing on algorithmic designs, TESRAC offers a flexible, extensible evaluation architecture that integrates various reduction and prioritization tools, enabling multi-criteria decision-making based on reduction size, coverage, and execution time. By repurposing existing test case prioritization tools (e.g., Kanonizo) and reduction tools (e.g., EvoSuite, Randoop, Testler), TESRAC not only benchmarks performance across diverse scenarios but also demonstrates that prioritization techniques can be effectively adapted for test minimization purposes. Importantly, TESRAC supports fine-grained customization of decision criteria, making it particularly suitable for ROSAS, where balancing testing time, resource constraints, and safety coverage is critical. In such systems, which often involve large and highly redundant test suites generated from multi-modal scenarios, TESRAC’s ability to automatically compare, tune, and select the optimal reduction strategy offers significant potential to improve regression testing efficiency while ensuring essential behavior coverage.

\subsubsection{Discussion and Remarks}
Test suite minimization remains a crucial strategy for enhancing the efficiency and scalability of testing processes, particularly in complex, resource-constrained environments such as ROS-based autonomous systems. While traditional TSM techniques such as greedy algorithms, clustering-based approaches, and coverage-driven subset selection provide a strong foundation, their direct application to ROSAS poses significant challenges.

The heterogeneous, dynamic, and asynchronous nature of ROSAS demands minimization strategies that go beyond syntactic or structural coverage preservation. Effective TSM for ROSAS must incorporate semantic understanding of multi-modal sensor interactions, temporal dependencies arising from asynchronous node communication, and cross-module functional behaviors. Preserving diversity across environmental scenarios, sensor conditions, and system responses is essential to maintain meaningful fault detection capability within minimized test suites.

Although existing studies in cyber-physical systems and autonomous driving domains offer promising methodologies such as semantic clustering of scenarios and graph-based dependency analysis, there remains a clear research gap in adapting these techniques to the unique requirements of ROS architectures. Future directions should emphasize context-aware, behavior-preserving minimization strategies that balance efficiency with safety-critical system validation needs.

\subsection{Test Case Selection}
Test case selection focuses on selecting a subset of existing test cases that are relevant to recent changes in a system, aiming to optimize regression testing efforts without compromising fault detection effectiveness. As summarized in Section \ref{tech_traditional}, TCS plays a vital role in traditional software engineering by reducing unnecessary test executions, thus improving testing efficiency and accelerating development cycles.

In the context of ROSAS, the importance of TCS becomes even more pronounced. ROSAS are typically modular, distributed, and highly dynamic, involving complex interactions across perception, planning, control, and actuation modules. Incremental updates to a single node, sensor driver, or topic communication protocol can propagate through the system and affect behaviors in subtle and non-obvious ways.

However, applying traditional TCS strategies directly to ROSAS environments presents unique challenges. Unlike monolithic software systems, ROSAS exhibit asynchronous communication patterns, multi-modal sensor fusion, and real-time decision-making processes, which complicate the identification of affected components and relevant test cases. Moreover, environmental variability and sensor-induced uncertainty introduce additional layers of complexity in determining which tests are truly impacted by a given system change.

Effective TCS in ROSAS must therefore combine structural analysis with behavioral, environmental, and runtime dependency considerations, ensuring that selected test subsets remain sufficient to validate system correctness and safety-critical functionalities.

\subsubsection{Categories of TCS Techniques}
Several categories of test case selection techniques have been developed to improve the efficiency of regression testing by identifying and executing only those tests that are impacted by system changes. When applied to ROSAS, these techniques must be carefully adapted to handle the modular, asynchronous, and sensor-driven nature of the systems. This section summarizes the primary TCS categories and their relevance to ROSAS environments.

\paragraph{\textbf{Static Analysis-based Selection}}
Static analysis-based TCS techniques rely on analyzing the program’s source code, system architecture, or configuration files to determine dependencies between components and to identify which test cases are affected by a change. This method is widely used in traditional software systems due to its scalability and low runtime overhead \cite{liu2023more, mafi2024regression, kauhanen2021regression, lou2024automated, xu2021requirement}.


In ROSAS, static analysis can be extended to: (1) semantic static analysis: enhancing the precision of TCS by identifying code changes with semantic impact \cite{liu2023more, mafi2024regression}; and (2) multi-dimensional static analysis: comprehensively assessing the impact paths of code changes through the integration of various static analysis techniques (such as build dependency analysis) \cite{Timperley2024ROBUST2B}. However, static analysis alone may miss runtime behaviors that emerge from dynamic environment interactions, making it insufficient for fully capturing the impact of changes in ROSAS.

Liu et al. \cite{liu2023more} proposed a regression TCS method based on semantic modification change reasoning. The core idea is to identify parts of code changes that may affect program semantics through static analysis, thereby enhancing the accuracy of TCS. In this method, the researchers first systematically classified various types of code changes, summarizing 29 different change types and further merging them into 13 “findings”, where 11 were explicitly defined as “semantic modification” changes, i.e., those types of changes that may alter the program's execution behavior or output results. On this basis, the authors applied these semantically sensitive change patterns to the improvement of two classic regression TCS techniques, Ekstazi and Neos. By more finely delineating the scope of change impact, they enhanced these two techniques' ability to identify affected test cases, thereby improving the effectiveness and efficiency of testing. Mafi et al. \cite{mafi2024regression}, in the context of test-driven development (TDD), proposed a regression TCS method that integrates static analysis and natural language processing (NLP) techniques. The core idea of this method is to establish structural dependency relationships between test cases and source code through static analysis, thereby identifying the code units covered by test cases. Meanwhile, NLP techniques are utilized to conduct semantic analysis on the descriptive text of test cases, extracting their functional intentions and behavioral characteristics. On this basis, the researchers combined code-level dependency information with the semantic information of test cases to more precisely determine the set of test cases that may be affected by code changes. This method not only improved the precision of traditional dependency-based TCS but also introduced semantic-level understanding, enhancing the ability to identify the scope of change impact. Thus, it improved the efficiency and intelligence level of regression testing while ensuring testing effectiveness.

Lou et al. \cite{lou2024automated} collected two million code patches in their research and proposed a TCS method based on static analysis. This method analyzes the static characteristics of patches to predict their impact on program behavior and selects the most relevant test cases for verification accordingly. Specifically, the method includes three key steps: (1) analyze the patch content to identify the scope and type of code changes; (2) determine the code elements that may be affected based on these changes; and (3) select the corresponding test cases according to the affected code elements. By combining three static analysis techniques, change impact analysis, data flow analysis, and control flow analysis, this method achieves efficient and precise TCS.
Elsner et al. \cite{elsner2022build} proposed a Build System Aware Multi-language Regression Test Selection technique, designed to address the challenges of large, heterogeneous, and cross-language codebases within CI pipelines. Unlike conventional regression TCS methods that focus on single language or source code only dependencies, their approach integrates static and dynamic analysis to capture per-test dependencies across languages and non-code artifacts, including configuration files and dynamically linked libraries. Through the combination of system call tracing, JVM class loader monitoring, and build dependency analysis, the method enables precise selection of both impacted tests and the corresponding build modules. Empirical results demonstrated its ability to safely exclude up to 75\% of tests on average, while reducing end-to-end CI pipeline times by up to 63\% without missing real faults. This technique is highly applicable to ROSAS, which feature similarly complex, multi-language components, cross-module interactions, and configuration-driven behaviors. In such systems, where testing every scenario is impractical due to limited resources and frequent software updates, build system aware and cross-artifact regression TCS techniques offer a scalable and safe solution for reducing redundant test executions while preserving fault detection effectiveness.

\paragraph{\textbf{Dynamic Analysis-based Selection}}
Dynamic analysis-based TCS techniques utilize runtime information, such as execution traces, sensor logs, or coverage data, to determine which test cases exercised changed components during actual system executions. These methods can capture data and control flow dependencies that are difficult to infer statically \cite{huang2024neuron, gao2022adaptive, ibias2021using, hao2023test, guizzo2024speeding, dobslaw2023generic, arrieta2023some}.


For ROSAS, dynamic analysis can involve: (1) behavioral feature analysis: involves monitoring and analyzing behavioral features such as neuron activation patterns, gradient changes, and critical execution paths; and (2) execution trace and dependency analysis: based on execution traces, function call paths, and input-output dependencies observed during runtime, this technique constructs a dynamic correlation model that captures the relationships between test cases and system behavior changes. Dynamic techniques are better suited to handle the non-deterministic and environment-sensitive behaviors characteristic of ROSAS. However, they often incur higher runtime costs and require extensive trace collection infrastructure.

Huang et al. \cite{huang2024neuron} proposed a TCS method for deep neural networks based on neuron sensitivity. This method evaluates the sensitivity of neurons to inputs through dynamic analysis techniques, identifies neurons that have a critical impact on model outputs, and screens test cases that can effectively cover these key neurons, thereby enhancing the efficiency of TCS and the ability to detect faults. Its innovation lies in using neuron sensitivity as the basis for TCS, which enhances the observability of the model's internal behavior during the testing process. However, this method has certain computational overhead in sensitivity calculation and is limited in its adaptability to different neural network architectures, requiring further optimization and expansion. Gao et al. \cite{gao2022adaptive} proposed a dynamic TCS method for deep neural networks. This method captures subtle changes in model behavior by monitoring the activation patterns and gradient changes of neurons in real time during the testing process, thereby adaptively screening the most representative test cases. The method can more accurately reflect the operating state of the model under different inputs, improving test coverage and fault detection capabilities, and has strong targeting and effectiveness.

Ibias et al. \cite{ibias2021using} proposed a TCS method based on mutual information theory, aiming to improve test efficiency and diagnostic capabilities. This method quantifies the mutual information relationship between test cases and system faults to build a dynamic selection model, thereby screening the most informative and diagnostically valuable subset from a large-scale test case set. While measuring the correlation between test cases and faults, the researchers introduced the “biased mutual information” metric, which also considers the redundancy among cases, effectively enhancing the diversity and execution efficiency of the test set. Arrieta et al. \cite{arrieta2023some} proposed a search-based TCS method based on a seeding strategy, aiming to improve test efficiency and defect detection capabilities. This method identifies representative and high-value seed test cases through dynamic analysis of the test execution process and constructs an efficient TCS framework. The researchers designed a dynamic seed selection mechanism based on evolutionary algorithms, which can adaptively adjust the selection and generation of seed cases according to test objectives and system characteristics. This strategy combines historical execution information with real-time feedback to optimize the diversity and directionality of the seed set, thereby improving test coverage and fault detection efficiency under limited resources. Elsner et al. \cite{elsner2023binaryrts} proposed BinaryRTS, a dynamic regression TCS technique designed to address the challenges of cross-language and binary-level dependencies in modern software systems. Unlike traditional approaches that rely on source-level or file-level analysis, BinaryRTS uses dynamic binary instrumentation to capture per-test execution traces of covered functions and accessed external files, including those in compiled C++ binaries and non-code artifacts. This enables precise and safe selection of impacted tests across heterogeneous codebases and languages. Empirical results from a large-scale industrial deployment demonstrated that BinaryRTS can safely exclude up to 74\% of tests without missing real faults, significantly reducing CI pipeline execution times. For ROSAS, which often integrate C++ modules, dynamic libraries, and multi-language components with sensor-driven and configuration-based behaviors, BinaryRTS offers clear advantages. Its binary-level, language-agnostic, and artifact-aware design makes it particularly well-suited for selecting tests affected by software updates, sensor drivers, middleware changes, and configuration alterations, all of which are common and critical in autonomous system deployments. As such, BinaryRTS represents a scalable and practical solution to optimize test execution while ensuring reliability in complex and evolving autonomous environments.

\paragraph{\textbf{Dependency Graph-based Selection}}
Dependency graph-based TCS approaches build explicit graphs representing dependencies among system components, and then select test cases based on the transitive closure of affected nodes or modules following a change \cite{d2022catto}.


In ROSAS, dependency graphs are utilized to represent the structure and behavior of software, thereby enabling the design of various TCS strategies based on these graph models. Dependency graph-based selection offers a promising middle ground between purely static and purely dynamic methods. It enables a more comprehensive understanding of change impact while maintaining acceptable computational overhead, making it particularly attractive for large, modular ROSAS.

CATTO \cite{d2022catto} selects test cases related to code changes by constructing and comparing call graphs before and after software modifications. The core of this method lies in leveraging the structural information of call graphs to identify execution paths that may be affected by code modifications, thereby precisely selecting test cases that cover these paths. Compared with traditional methods based on text differences or data flow analysis, the graph-based approach can more intuitively reflect the execution logic of programs and the dependencies between code segments, thereby more effectively reducing the number of test cases while maintaining a high fault-detection capability. GameRTS \cite{yu2023gamerts} proposes a novel regression TCS framework specifically tailored for video game software, which shares important similarities with autonomous systems, such as complex states, high-frequency updates, and long-running tests. Traditional techniques are impractical in such contexts due to the size and state-space explosion problems. To address this, GameRTS models test cases as state transition graphs (STGs) and performs fine-grained change impact analysis across code, design flowcharts, and resource files to identify affected states and actions. By leveraging the Markov property of gameplay, it selects minimal test traces that still achieve high fault detection capability while significantly reducing test costs. This approach ensures practicality without sacrificing safety. Importantly, this STG-based method is highly applicable to ROSAS, where software similarly integrates rule-based and learning-based modules, and runtime behaviors depend on complex environment states. The idea of modeling system behavior as abstract state transitions and performing selective re-testing can help reduce the regression testing burden while maintaining coverage and safety guarantees in autonomous system updates.

\paragraph{\textbf{Data-driven Selection}}
Unlike traditional static and dynamic analysis methods that rely primarily on source code structure or execution traces, data-driven approaches utilize a wide range of empirical data and statistical models to guide test selection decisions. These techniques leverage historical execution data, system logs, performance metrics, and even latent representations to predict which test cases are most likely to be affected by code changes or which are most valuable for regression verification \cite{sun2023robust, demir2024test, liang2023late, birchler2023cost, li2021conditional, bao2023defense, zhang2023optimizing, birchler2023teaser}.

In the context of ROSAS, data-driven selection offers particularly strong potential to address key testing challenges: (1) handling non-determinism and runtime variability: data-driven models can learn from historical test results and runtime observations to better predict which test cases are sensitive to recent changes; (2) capturing semantic and context-dependent impacts: by analyzing multi-source runtime data (e.g., sensor logs, decision outputs), data-driven selection can identify test cases that exercised semantically similar or critical scenarios; and (3) supporting continuous learning and test optimization: as autonomous systems are updated and refined, data-driven models can incrementally learn from new data, improving test selection precision without manual intervention. For ROSAS, data-driven test selection can encompass the following aspects: (1) uncertainty-based test selection: guiding the selection of test cases by quantifying the predictive uncertainty of input data;  and (2) historical behavior patterns: extracting patterns from historical continuous integration data, build metadata, and test execution records for test selection.

RTS \cite{sun2023robust} is a TCS method that integrates metamodels with uncertainty metrics, aiming to enhance the efficiency and effectiveness of data selection for deep learning model testing. This method quantifies the model's predictive uncertainty for input data by calculating various uncertainty metrics of the test data (such as minimum confidence, entropy, margin distance, and deep ensemble variation scores), thereby providing a data-driven basis for TCS. Moreover, the method introduces a metamodel to learn the mapping between uncertainty metrics and model prediction errors, thereby optimizing the screening process of test cases. The construction of the metamodel is entirely based on features extracted from test data, while also incorporating out-of-distribution (OOD) detection techniques to identify samples that deviate from the training data distribution, enhancing the sensitivity and adaptability of test selection to anomalous inputs. Overall, this method not only improves testing efficiency but also enhances the ability to detect potential model defects.

SDC-Scissor \cite{birchler2023cost} is a data-driven TCS method based on machine learning, aimed at optimizing the simulation testing efficiency of self-driving car software. This method constructs a predictive model by extracting key features (such as road properties and driving path statistics) from a large number of simulation test cases to identify test cases that are likely to reveal software failures. Based on the model's prediction results, SDC-Scissor can effectively filter out ``informationless" test cases with high execution costs but low detection value, thereby significantly reducing testing costs and resource consumption. HybridCISave \cite{jin2023hybridcisave} exemplifies the strength of data-driven selection techniques by leveraging historical CI data, build metadata, and test execution records to guide both build and TCS decisions. Its core mechanism integrates multiple data-driven predictors such as test failure history, file-level change impact, and previous build outcomes into a supervised learning model, random forest. This model predicts which builds and associated tests can be safely skipped without risking undetected regressions. Additionally, HybridTestSkip uses ensemble-like consensus voting among different historical predictors to make test selection decisions based on empirical data patterns rather than static rules. This multi-source, statistically-driven decision-making process highlights the data-driven nature of HybridCISave, where selection relies on learned patterns extracted from rich operational and testing datasets. For ROSAS, HybridCISave's data-driven methodology is highly relevant and adaptable. Autonomous systems generate large volumes of runtime data, such as perception logs, planning traces, and control signals, which can be harnessed similarly to CI data. By training data-driven models on this information, HybridCISave's approach could support intelligent regression test selection, prioritizing tests for modules where historical patterns indicate higher risk, or skipping tests in stable modules with minimal recent changes. Furthermore, ensemble and learning-based decision strategies, as used in HybridTestSkip, can help handle the uncertainty and non-determinism common in autonomous systems, making selection decisions more robust and context-aware.

\subsubsection{Discussion and Remarks}
Test case selection is a fundamental strategy for improving regression testing efficiency by selecting a relevant subset of test cases that sufficiently validate recent system changes. In traditional software engineering, based on static analysis, dynamic execution tracing, and dependency modeling, TCS techniques have achieved significant success in reducing testing costs and accelerating development cycles.

However, applying TCS to ROS-based autonomous systems introduces distinctive challenges. The modular, asynchronous, and environment-driven nature of ROSAS complicates the identification of impacted components and the propagation of changes through the system. Dynamic node dependencies, non-deterministic message exchanges, multi-modal sensor interactions, and transitive behavioral impacts must all be considered to ensure comprehensive yet efficient test selection.

Existing studies in cyber-physical systems, autonomous driving, and runtime behavior-driven testing provide valuable methodologies for adapting TCS to ROSAS. Nonetheless, current approaches often fall short in fully capturing the complexities of ROS architectures, particularly in handling asynchronous communication patterns and environmental variability.

Future research directions should focus on developing hybrid TCS frameworks that integrate static structural analysis with dynamic runtime behavior profiling and environmental context modeling. Leveraging graph-based representations of node-topic-service interactions, sensor influence networks, and multi-modal behavior traces could enable more precise and safety-aware test selection strategies tailored for the unique demands of ROS-based autonomous systems.

\subsection{Limitations of Emerging Techniques for ROSAS}

Despite the emergence of new techniques in TCP, TSM, and TCS that have shown potential in specific scenarios, they still fall short when addressing the unique challenges posed by ROS-based Autonomous Systems. It is worth noting that most of these emerging techniques focus primarily on the testing optimization of deep learning models rather than directly targeting the characteristics of ROSAS.

For instance, many emerging methods leverage the internal behaviors of deep learning models, such as neuron activation patterns, to guide the selection and prioritization of test cases. However, these approaches often overlook the multi-modal sensor data, asynchronous communication patterns, and real-time requirements inherent in ROSAS. Moreover, the complexity of ROSAS lies in their distributed architecture and dynamic interaction with the environment, characteristics that render both traditional and emerging testing optimization techniques difficult to apply directly.

Specifically, existing testing optimization techniques fall short in the following aspects:

\begin{itemize}

\item \textbf{Complexity of Multi-modal Data:} ROSAS need to process high-dimensional, multi-modal data from various sensors, such as cameras, LiDAR, and radar. Most existing testing optimization techniques focus on single-modal data and struggle to effectively handle the complexity of multi-modal data.

\item \textbf{Dynamic and Asynchronous Behavior:} The dynamic nature and asynchronous communication patterns of ROSAS complicate the selection and prioritization of test cases. Existing techniques often assume deterministic system behavior, which does not align with the actual operation of ROSAS.

\item \textbf{Real-time and Safety Requirements:} ROSAS typically operate under strict time constraints while ensuring system safety. Existing testing optimization techniques generally lack direct support for real-time and safety requirements.

\end{itemize}

These shortcomings indicate that existing testing optimization techniques, whether traditional or emerging, still have significant limitations when dealing with the specific issues of ROSAS. Therefore, it is necessary to explore new methods and strategies to better address the testing challenges in ROSAS. Next section will delve into these limitations and propose future research directions and potential solutions.

\section{Challenges and Research Opportunities for ROS-based Autonomous Systems} \label{sec:challenge}
As autonomous systems are increasingly integrated into real-world environments, ensuring their safety, reliability, and robustness through effective regression testing has become critically important yet exceptionally challenging. Traditional regression test optimization techniques such as regression test case prioritization, minimization, and selection have proven highly successful in conventional software engineering. However, when directly applied to ROSAS, these methods face significant limitations due to fundamental differences in system architecture, behavior, and operational context.

ROS-based autonomous systems differ markedly from traditional software systems, primarily due to their intrinsic complexity and highly dynamic interactions with unpredictable environments. They typically incorporate advanced modules for perception, prediction, planning, and control, combining stochastic machine learning models with deterministic rule-based logic. Consequently, the behavior of ROSAS often exhibits non-deterministic, context-dependent, and complex interactions that pose considerable difficulties for traditional testing methodologies.

In this chapter, we systematically explore these distinctive challenges, addressing Research Question 2: ``What are the limitations and gaps in current optimization methods when applied to ROS-based autonomous systems?" Specifically, we analyze how issues such as inadequacies of traditional coverage metrics, complexities arising from multi-modal and high-dimensional sensor data, dynamic and non-deterministic system behaviors, and the absence of reliable test oracles complicate effective regression testing optimization. This analysis draws upon a substantial body of existing literature to underline the practical implications and theoretical constraints faced by existing approaches.

Subsequently, we synthesize critical insights distilled from existing research, highlighting valuable lessons learned regarding the effectiveness and limitations of current methods. These insights include the importance of semantic-aware coverage metrics, the benefits of integrating hybrid optimization strategies, the advantages of neurosymbolic approaches for enhanced explainability, and the adaptability provided by data-driven selection methods. Collectively, these insights form the foundation for understanding how current research informs the practical challenges of regression testing for ROSAS.

Finally, guided by these insights and recognizing current limitations, this chapter proposes actionable research directions addressing Research Question 3: ``What are the promising future directions and emerging techniques for regression testing optimization in ROS-based autonomous systems?" We outline several innovative approaches and novel techniques specifically designed to confront the complexities associated with ROSAS testing. These future directions encompass semantic-aware coverage metric enhancements, exploitation of multi-source foundation models, further advancement of neurosymbolic techniques, adaptive and real-time optimization frameworks, and robust oracle development via scenario-based and metamorphic testing approaches.

By clearly identifying these challenges, synthesizing key research insights, and proposing a forward-looking roadmap, this chapter provides a comprehensive foundation for advancing regression testing optimization methods, ultimately facilitating safer, more robust, and scalable deployment of autonomous systems in real-world operational environments.

\subsection{Current Challenges in Regression Testing for ROSAS}
Testing ROS-based autonomous systems introduces a wide array of unique challenges that fundamentally differentiate it from conventional software testing. This section details the key obstacles that hinder the effective validation, verification, and optimization of regression testing in ROSAS environments.

\subsubsection{Coverage Metrics Inadequacy in ROSAS Contexts}
Traditional coverage metrics such as statement coverage, branch coverage, and neuron coverage have been the foundation of test optimization techniques in conventional software engineering. These metrics serve as proxies for assessing the thoroughness of testing by tracking which parts of the system have been exercised. Numerous regression testing strategies adopt these metrics to guide test case prioritization and minimization. For instance, coverage-guided approaches such as the dominating set-based method \cite{demir2022dominating} and search-based strategies like GASSER \cite{coviello2022gasser} leverage structural or execution-based coverage to optimize test suites. Similarly, neural coverage has been proposed in pattern-based prioritization frameworks \cite{harel2020neuron} to evaluate how much of a neural network’s latent space is stimulated by test cases.

However, these traditional coverage metrics fall short when applied to ROSAS, which integrate heterogeneous modules including rule-based planners, probabilistic predictors, and deep learning-based perception models. As highlighted in multiple studies, the assumption that structural coverage correlates with behavioral completeness no longer holds in such hybrid architectures. For example, in the test prioritization work \cite{deng2022scenario}, although structural diversity in simulated driving scenarios was achieved, the correlation between code-level coverage and critical behavioral faults remained weak. Likewise, the search-based selection and prioritization framework for autonomous driving systems \cite{lu2021search} adopts scenario-based coverage rather than relying solely on code coverage, demonstrating a shift in priority toward semantic diversity.

Another critical limitation is that structural metrics often neglect the semantic meaning of high-dimensional, multi-modal inputs like LiDAR point clouds or camera images, which are the primary data sources in autonomous systems. Segment-based prioritization \cite{huynh2024segment} and ATM \cite{pan2023atm} illustrate how input similarity and data evolution can be more predictive of test value than low-level statement coverage. Moreover, in neuron-based prioritization \cite{yan2022test}, despite using deep model internals, the study finds limited effectiveness in identifying behaviorally diverse or semantically novel cases, especially in high-dimensional input spaces typical in perception modules.

Importantly, many studies indirectly expose the inadequacy of traditional metrics by proposing alternatives. For instance, scenario-based reduction frameworks \cite{deng2022scenario} and multi-objective formulations using semantic features \cite{xue2020multi, coviello2022gasser} prioritize test scenarios based on behavioral diversity, safety criticality, or operational context relevance, not merely structural code paths. These approaches implicitly acknowledge that code coverage metrics do not align with the real-world safety and functional goals of autonomous systems.

In ROSAS, sensor data frames are streamed at millisecond or nanosecond intervals, leading to enormous volumes of high-dimensional input. Metrics such as statement or neuron coverage cannot distinguish between semantically identical frames (e.g., similar weather or traffic scenes) and diverse ones that may trigger rare edge behaviors. This inadequacy highlights the need for semantic-aware and frame-level vectorization strategies, which are better suited to measure coverage across perceptual, planning, and control dimensions in ROSAS.

Therefore, while traditional coverage metrics remain useful for evaluating certain software properties, they are fundamentally insufficient for representing the behavioral richness, semantic diversity, and operational safety critical to autonomous system testing. Future work must explore vector-based, task-aware, and risk-sensitive coverage models that are contextually aligned with the high-stakes nature of ROS-based regression testing.



\subsubsection{Complexity of Multi-modal and High-dimensional Data}
ROS-based autonomous systems operate in dynamic environments, relying on a diverse array of sensors such as LiDAR, radar, cameras, GPS, and inertial measurement units (IMUs) to perceive and interpret their surroundings. These sensors generate vast amounts of high-dimensional data, often in the form of point clouds, images, and time-series signals, which must be processed and fused in real-time to enable safe and effective decision-making. The heterogeneity and volume of this data introduce significant challenges for regression testing, particularly in terms of test case prioritization and coverage assessment.

Traditional coverage metrics, such as statement or branch coverage, are insufficient for capturing the nuanced behaviors elicited by complex sensor inputs \cite{strickland2018deep}. For instance, minor variations in sensor data such as a slight change in lighting conditions or the presence of a new object can lead to significant differences in system behavior, even if the underlying code paths remain unchanged. This disconnect underscores the need for testing approaches that consider the semantic content of sensor data and its impact on system performance.

To address these challenges, researchers have proposed scenario-based testing frameworks that focus on the semantic features of driving environments. Deng et al. \cite{deng2022scenario} introduced STRaP (Scenario-based Test Reduction and Prioritization), a framework that encodes driving scenes into feature vectors based on a predefined schema encompassing elements like traffic lights, stop signs, and pedestrian crossings. By segmenting driving recordings into semantically distinct scenarios and prioritizing them based on coverage and rarity, STRaP effectively reduces test suite size while maintaining fault detection capabilities.

Furthermore, the fusion of multi-modal sensor data necessitates sophisticated synchronization and calibration techniques. Luo et al. \cite{luo2025accidents} proposed a comprehensive method for generating multi-modal perception data by integrating image data, point clouds, and IMU/GPS readings within the ROS framework. This approach facilitates accurate visualization and analysis of the vehicle's surroundings and behavior, enhancing the evaluation of autonomous driving systems.

The complexity of multi-modal data also extends to anomaly detection. Noorani et al. \cite{noorani2024multimodal} developed a robust anomaly detection system for autonomous cyber-physical systems by integrating multiple sensor modalities, including LiDAR, odometry, and network traffic data. Their approach leverages a vector-based reconstruction loss function, significantly improving the detection of subtle anomalies by preserving the contributions of individual features.

In summary, the high-dimensional and multi-modal nature of sensor data in ROSAS presents unique challenges for regression testing. Traditional coverage metrics fall short in capturing the semantic richness of sensor inputs and their influence on system behavior. Emerging approaches that incorporate semantic scene understanding, multi-modal data fusion, and advanced anomaly detection techniques offer promising avenues for enhancing the effectiveness and efficiency of regression testing in autonomous systems.

\subsubsection{Dynamic and Non-deterministic System Behavior}
ROS-based autonomous systems operate in complex, dynamic environments where unpredictability is inherent. These systems must adapt to varying conditions, such as fluctuating sensor inputs, changing environmental factors, and evolving internal states. This dynamism introduces non-deterministic behavior, where identical inputs may not consistently produce the same outputs, posing significant challenges for regression testing.

Non-determinism in autonomous systems arises from multiple sources. Learning-based components, such as those employing reinforcement learning algorithms, can evolve their behavior over time, leading to different responses to the same stimuli. For instance, a study \cite{kashyap2025autonomous} on autonomous navigation using ROS2-based TurtleBot3 demonstrated that reinforcement learning algorithms like TD3 and DDPG enable the robot to adapt its navigation strategies in both static and dynamic environments, resulting in behavior that changes as the system learns from new experiences.

Moreover, the integration of concurrent processes and asynchronous communication in ROS-based systems contributes to non-deterministic outcomes. The timing and order of message passing between nodes can vary \cite{TomLooman}, leading to different execution paths and system states. This variability complicates the testing process, as traditional deterministic testing approaches may fail to capture the full spectrum of possible behaviors.

To address these challenges, researchers have proposed various strategies. Runtime verification and field-based testing have emerged as effective methods for monitoring system behavior in real-world scenarios. A recent study \cite{jiang2025step} provided guidelines for implementing these techniques in ROS-based drone systems, emphasizing the importance of designing systems that facilitate observability and controllability during operation.

Additionally, simulation-based testing frameworks have been developed to model and analyze the dynamic behavior of autonomous systems. These frameworks allow for the systematic exploration of different scenarios and the identification of edge cases that may lead to unexpected behavior. For example, a rigorous simulation-based testing method \cite{li2024rigorous} was proposed to decompose complex scenarios into simpler ones, enabling the analysis of autopilot decisions in dynamic environments.

In conclusion, the dynamic and non-deterministic nature of ROSAS necessitates advanced testing methodologies that can accommodate variability and unpredictability. By leveraging runtime verification, field-based testing, and sophisticated simulation frameworks, developers can better assess system behavior, identify potential issues, and enhance the reliability and safety of autonomous systems.

\subsubsection{Lack of Reliable Test Oracles}
In the context of ROS-based autonomous systems, establishing reliable test oracles, which are mechanisms that determine the correctness of system outputs given specific inputs, is a significant challenge. The complexity arises from the systems' integration of diverse components, including perception modules, decision-making algorithms, and control systems, all operating in dynamic and unpredictable environments. Traditional test oracles, which often rely on predefined expected outputs, are insufficient for such systems due to the difficulty in specifying correct behavior for every possible scenario.

The ``oracle problem" in software testing refers to the challenge of determining whether the output of a system under test is correct for a given input \cite{pezze2014automated}. This problem is exacerbated in autonomous systems, where the expected behavior may not be well-defined or may vary depending on context. As noted in the literature \cite{duque2023towards}, the absence of reliable test oracles hinders the effectiveness of regression testing, as it becomes challenging to detect unintended changes in system behavior.

To address this issue, researchers have explored various approaches. One method involves using object state data to derive test oracles \cite{duque2023towards}. By monitoring the state of objects during test execution and comparing them across different system versions, it is possible to detect behavioral changes that may indicate faults. This approach enhances the detection of unintended behavior changes, providing a more effective means of regression testing in the absence of explicit expected outputs.

Another promising avenue is the application of machine learning techniques to generate test oracles. For instance, the DARIO \cite{arrieta2021using} system employs regression learning algorithms trained on data from previously tested versions to predict the quality of service of elevator dispatching algorithms. This method allows for the automatic validation of system behavior, even in the absence of formal specifications, by learning from historical data and identifying deviations from learned patterns.

Furthermore, metamorphic testing has been proposed as a means to alleviate the oracle problem \cite{fontes2021using}. This technique involves identifying metamorphic relations, which are properties of the system that should hold true across multiple inputs and outputs, and using them to validate system behavior. By focusing on the relationships between inputs and outputs rather than specific expected outputs, metamorphic testing provides a way to detect faults in systems where traditional test oracles are unavailable or impractical \cite{deng2022declarative}.

Despite these advancements, the development of reliable test oracles for ROSAS remains an open research area. The inherent complexity and variability of autonomous systems necessitate continued exploration of innovative approaches to effectively validate system behavior and ensure reliability in real-world applications.

\subsection{Research Insights from Existing Optimization Methods}
While traditional test optimization strategies face significant limitations in ROS-based autonomous systems, emerging research directions offer promising avenues to overcome these challenges. Based on the unique characteristics of ROSAS and the analysis presented in this survey, we identify several key research insights that can guide future advancements in autonomous system testing.

\subsubsection{Semantic Representation Matters}
Emerging evidence highlights that incorporating semantic and scenario-based context, rather than relying purely on syntactic or structural coverage, significantly enhances test effectiveness for ROS-based autonomous systems. Traditional coverage metrics often fail to capture behavioral correctness in systems with complex perception, planning, and decision-making logic. To bridge this semantic gap, recent studies have introduced novel techniques to encode, track, and prioritize tests based on their semantic relevance.

In the work \cite{deng2022scenario}, the authors propose a multi-step pipeline that extracts scenario-relevant information from autonomous vehicle simulations. By clustering simulation episodes and selecting representative trajectories based on scenario diversity, they achieve better behavioral coverage while minimizing the number of test cases. This approach is particularly effective for ROSAS, where rich contextual understanding (e.g., traffic configurations, object interactions) is often more important than low-level execution path coverage.

Li et al. \cite{li2024semantic} introduce SatTCP, a semantic-aware test prioritization method that computes embeddings of code and tests based on semantic similarity. Their two-phase framework first constructs vector representations using NLP techniques and then computes semantic distances between test cases and code changes. This embedding-based prioritization improves regression test selection for CI pipelines, especially when traditional structural analysis lacks precision. ROSAS, which often contain neural modules and data-driven decision points, can benefit from such semantically-informed test prioritization by focusing on meaningful functional shifts.

In \cite{liu2023more}, Liu et al. analyze real-world commits to identify 29 categories of semantics-modifying code changes. They enhance two regression test selection (RTS) tools (Ekstazi \cite{gligoric2015ekstazi} and STARTS \cite{legunsen2017starts}) to reason about such changes and selectively avoid re-running tests unaffected at a behavioral level. This precision-aware RTS method avoids false positives from surface-level change analysis and speeds up CI cycles. While their method is designed for general Java applications, its precision-enhancing philosophy is applicable to ROSAS where behavior changes may not always be reflected in syntactic deltas.

Finally, in \cite{zheng2024testing}, Zheng et al. propose using LLMs to generate scenario-level test cases based on natural language specifications, past test logs, and operational constraints. This approach shifts test generation and selection from code-centric to intent-driven, enabling validation of high-level semantics, especially in systems with perception and planning components. Their method demonstrates strong potential for testing ROSAS, where corner-case scenarios are not easily captured by structural metrics alone.

Together, these studies underscore the growing importance of semantic-aware techniques in regression testing optimization. They point toward a shift from conventional structural metrics to representations that encode functional intent, behavioral diversity, and contextual correctness, which are attributes essential for the safety and robustness of ROS-based autonomous systems.

\subsubsection{Hybrid Approaches Are Essential}
Recent studies suggest that hybrid approaches, which integrate static, dynamic, and data-driven techniques, offer superior adaptability and effectiveness over singular methods in test case selection and prioritization. These strategies are especially critical in the context of ROS-based autonomous systems, which exhibit non-determinism, dynamic feedback loops, and high-volume multi-modal sensor input. Hybrid frameworks can harness different forms of information (e.g., historical test results, dependency structures, code changes) to achieve more robust and context-aware regression testing.

HybridCISave \cite{jin2023hybridcisave} exemplifies such a hybrid methodology by integrating both build-level and test-level data to improve continuous integration pipelines. By combining commit dependency analysis with dynamic test execution results, HybridCISave identifies both relevant builds and their associated test cases to prioritize and select. This combined perspective yields increased cost efficiency and faster feedback, while minimizing the risk of skipping fault-revealing test cases. Its ability to correlate build changes with test outcomes allows it to outperform single-focus techniques, particularly under time-constrained CI settings.

RoughTCP \cite{guaceanu2024leveraging}, introduced as a clustering-based unsupervised technique, leverages rough set theory to organize test cases without requiring labeled training data. By dynamically clustering test cases based on intrinsic patterns (e.g., historical faults, execution duration, frequency), RoughTCP can infer priority rankings that adapt to the volatile characteristics of real-world CI environments. This hybrid clustering and statistical method improves adaptability to uncertain or sparse data, common in ROSAS test pipelines. Moreover, RoughTCP’s non-reliance on supervised learning models makes it suitable for evolving systems where retraining is infeasible or unreliable.

CIBench \cite{jin2021cibench} serves as an empirical validation platform that aggregates and evaluates ten CI optimization techniques, including both selection and prioritization methods, across key dimensions such as time-to-feedback, computational cost, and failure detection rate. By reproducing and benchmarking diverse strategies under unified conditions, CIBench reveals how hybrid techniques that span static analysis, commit mining, and test outcome prediction tend to achieve the most balanced trade-offs. Its findings reinforce the necessity of blending approaches to accommodate the intricate dependencies and uncertainties characteristic of large-scale, continuously evolving systems such as ROSAS.

Collectively, these studies support the assertion that hybrid approaches are essential for addressing the multifaceted challenges of regression testing in ROSAS. Their ability to synthesize diverse evidence sources enables adaptive prioritization, maximizes fault detection under constrained resources, and ensures higher reliability in the testing of complex robotics software.

\subsubsection{Data-driven Techniques Improve Adaptability}
Data-driven techniques have emerged as critical components for improving the adaptability of test selection and prioritization processes in continuously evolving autonomous systems. These techniques leverage empirical features derived from system behavior, test outcomes, and code changes to guide test decisions dynamically. Compared to static or coverage-based methods, data-driven approaches are better suited to address the non-determinism, high-dimensionality, and hybrid architectures typical of ROSAS.

In \cite{sun2023robust}, the authors propose an RTS method that integrates neuron coverage features with an ensemble of data-driven learning models to rank and select test inputs capable of exposing incorrect DNN behaviors. Rather than relying solely on coverage percentage or gradient magnitudes, their learning-based test selector uses mutation sensitivity and output divergence patterns as supervision signals. This dynamic adaptation to network behavior ensures more effective fault exposure, especially in complex scenarios involving sensor fusion and decision boundary shifts. The approach demonstrates that static neuron coverage alone is insufficient, while feature-driven strategies provide resilience to network structure variations and unseen corner cases.

SDC-Scissor \cite{birchler2023cost} addresses the cost-effectiveness challenge of simulation-based testing in self-driving car software by using machine learning to predict which simulation tests are likely to be uninformative (i.e., not fault-revealing). Using road topology features (e.g., number of turns, cumulative angle, and segment lengths) extracted before test execution, the system classifies simulation tests as ``safe" or ``unsafe." These pre-execution features are learned from prior test outcomes, enabling the selection model to generalize to new scenarios and prioritize impactful test executions. The model achieved F1 scores up to 90\% and produced up to 170\% speedup over random baselines. This demonstrates the value of data-driven classification in adapting test effort to the behavior and topology of simulation inputs.

In \cite{zhang2023optimizing}, the authors propose DTPS, a dual-model framework that dynamically adjusts the proportion of tests to run per CI build based on predicted failure risk. DTPS combines a classifier to estimate the likelihood of build failure and a regressor to determine the necessary test proportion to reveal failures. The model is trained on build metadata and historical test execution logs, including code churn, file types changed, and project age. The approach avoids the binary decision-making of build-skipping techniques by preserving partial testing even under low predicted failure probability. DTPS improved fault detection by 19.9\% to 32.5\% compared to fixed-ratio test selection techniques, while maintaining cost-effectiveness, thereby showcasing the strength of data-driven proportional selection in dynamic environments.

\subsection{Future Research Directions}
This chapter outlines key future research directions that aim to overcome the identified challenges. These directions leverage emerging technologies such as multi-source foundation models, neurosymbolic reasoning, adaptive optimization strategies, and risk-aware testing frameworks. They collectively represent a roadmap for advancing the state of test optimization to better support the safe, robust, and scalable development of ROS-based autonomous systems.

\subsubsection{Advancing Semantic-aware Coverage Metrics} \label{frame-to-vector}
Traditional coverage metrics, such as statement or neuron coverage, are insufficient for evaluating the testing adequacy of ROSAS. Future studies should explore semantic-rich coverage metrics capable of representing diverse operational scenarios and behaviors comprehensively. To address the gap, we propose a frame-to-vector approach, where high-frequency sensor data frames (e.g., images, LiDAR scans) are vectorized into structured semantic representations. Given that the data frame generation frequency can reach nanosecond precision, frames can be captured and vectorized at almost any timestamp. Specifically, semantic information is extracted from sensor input data (e.g., camera images, 3D point clouds) and combined with system output labels (e.g., semantic segmentation bounding boxes, probability scores) to construct comprehensive vector representations. Through analyzing semantic proportions and similarities among these vectors, indirect measurements of test coverage can be achieved. This vector-based representation facilitates test optimization, enabling prioritization by analyzing semantic coverage percentages, minimization through redundancy elimination via vector similarity, and selective testing by identifying subsets of vectors with critical semantic content.




For example, STRaP \cite{deng2022scenario} has demonstrated the potential of encoding semantic features such as dynamic entities (vehicles, pedestrians, behaviors) and static entities (traffic lights, stop signs) into integer vectors. However, it relies heavily on manually defined scenario patterns, lacking flexibility to adapt to novel or unforeseen driving scenarios. Similarly, Neelofar et al. \cite{Neelofar2022IdentifyingAE} proposed an instance space analysis (ISA)-based method to identify safety-critical features in autonomous vehicle testing, but its results are constrained by the quality and diversity of existing test scenarios. These limitations highlight the need for more advanced semantic vector enrichment techniques that can dynamically capture a wider range of contextual information without manual intervention.



Current methods, such as STRaP, segment driving records into semantic fragments by comparing consecutive vector similarities and employ a smoothing mechanism via sliding windows \cite{deng2022scenario}. However, these approaches are limited in their ability to fully capture the temporal dynamics of complex behaviors. HIMap \cite{Zhou2024HIMapHR} addresses part of this challenge by integrating local positional information with overall shape and semantic data through its point-element interactor, but its focus on map reconstruction limits its efficiency for real-time temporal analysis. Future work should focus on developing more sophisticated temporal modeling techniques to ensure that coverage metrics can effectively represent the evolving nature of autonomous systems' behaviors.



Future frameworks should develop adaptive clustering methods that group vectors based on semantic and behavioral similarity, supporting scalable coverage estimation and guiding diversity-aware test prioritization and minimization. Dynamic clustering strategies can adjust based on operational feedback, ensuring continuous relevance to evolving system behaviors. Existing methods, such as TrafficGen \cite{feng2023trafficgen}, face challenges such as discontinuities when vehicles cross defined local coordinate boundaries, which can severely impair global consistency. PCPrior \cite{li2024test} attempts to address feature-based prioritization by extracting spatial and variability features from point clouds, but it struggles with diversity preservation. Adaptive clustering techniques could help mitigate such issues by dynamically adjusting to the spatial and temporal characteristics of the environment, thereby improving the accuracy and reliability of coverage metrics.



Frame-to-vector metrics should be integrated directly into test prioritization, selection, and reduction pipelines. By quantifying semantic novelty, risk exposure, and behavioral diversity, vector-based coverage models can guide optimization objectives more precisely than traditional code-centric metrics. STRaP \cite{deng2022scenario} prioritizes tests based on coverage and rarity, assigning higher weights to infrequent features to detect faults more effectively. However, its reliance on manual scenario definitions and simplistic vectorization restricts scalability and generalizability. CVNet \cite{Yan2024WhenVM} enhances vectorization coverage through its change-collector module and vector component learning model (VCLM), but its computational inefficiency limits real-time integration. VectorMapNet’s end-to-end pipeline \cite{Liu2022VectorMapNetEV} also offers insights into how vectorization can be embedded directly into system workflows without requiring extensive post-processing, thus reducing latency and increasing usability within automated testing environments. Future research should aim to develop more flexible and automated integration methods that can leverage the full potential of frame-to-vector metrics in test optimization pipelines.



Research should explore lightweight, scalable vectorization and similarity computation techniques to ensure that enhanced frame-to-vector coverage models are practical for large-scale ROSAS testing under real-time operational constraints. Existing methods, such as VAD \cite{jiang2023vad} and the ``Optimize \& Reduce" \cite{hirschorn2024optimize} method, face challenges in capturing irregular lane geometries and dynamic traffic behaviors. VectorMapNet \cite{Liu2022VectorMapNetEV} innovates with end-to-end vectorized HD map generation but suffers from feature space mismatches and hallucinations. These limitations underscore the need for more efficient and scalable vectorization techniques that can handle the computational demands of real-time testing while maintaining high accuracy and reliability.

\subsubsection{Enhancing Foundation Model-based Multi-modal Testing}
Future research can explore several directions for leveraging multi-source foundation models to enhance autonomous system testing, supported by recent literature on the effectiveness of such models. The advent of multi-modal foundation models such as CLIP \cite{radford2021learning}, LLaVA \cite{liu2023visual}, DeepSeek \cite{guo2025deepseek}, PaLM \cite{chowdhery2023palm}, and Gemini \cite{team2023gemini} offers unprecedented opportunities for autonomous system testing. These models possess strong generalization capabilities across diverse input modalities (text, vision, audio) and tasks. Unlike conventional deep learning models that require dedicated architectures and training regimes for each modality, foundation models inherently facilitate inductive reasoning and summarization across heterogeneous data sources, making them particularly suited for the complex multi-sensor environment found in ROS-based autonomous systems.


Foundation models can transform raw sensor data (e.g., camera images, LiDAR scans, radar returns) into rich semantic representations, enabling more meaningful analysis of environmental contexts, behavioral triggers, and scenario diversity. This moves beyond low-level pixel or point cloud comparisons, allowing for deeper semantic understanding. Recent studies demonstrate the efficacy of LLMs in integrating heterogeneous data, such as system logs, camera footage, and LiDAR signals, to capture contextual nuances. For example, Zheng et al. \cite{zheng2024testing} show that LLMs can automate the generation of realistic and diverse test scenarios by interpreting multi-modal sensor inputs. Similarly, Wang et al. \cite{Wang2023SoftwareTW} highlight the potential of LLMs in generating test inputs and debugging ROSAS, further validating their applicability in semantic abstraction for autonomous testing.


By interpreting heterogeneous sensor inputs simultaneously, foundation models facilitate automated scenario summarization, anomaly detection, and corner case identification. They can also generate synthetic yet realistic test scenarios by extrapolating from observed multi-modal patterns, enhancing test diversity. The effectiveness of multi-modal LLMs has been demonstrated in diverse applications. Huang et al. \cite{huang2023chatgpt} showcase their ability to integrate textual, audio, and image inputs for comprehensive diagnostics, suggesting similar potential in autonomous system testing. Additionally, SoVAR \cite{Guo2024SoVARBG} leverages multi-source foundation models to extract accident reports and generate generalized test scenarios, adapting to diverse map structures and environmental conditions. This approach significantly improves the automation and coverage of test case generation.


Semantic embeddings from foundation models can serve as proxies for environmental novelty, behavioral risk, or task relevance, informing dynamic test prioritization and selection strategies. This ensures that critical operational factors are adequately exercised during testing. TestGen-LLM \cite{Alshahwan2024AutomatedUT} exemplifies this capability by extending test suites and improving coverage through multi-source LLM integration, while employing rigorous filtering mechanisms to ensure reliability. Similarly, FlakyQ \cite{Rahman2024QuantizingLM} enhances test efficiency by predicting flaky tests, reducing resource waste, and supporting ROSAS testing through efficient feature extraction. These methods highlight how foundation models can optimize test selection and prioritization.


Pre-trained multi-modal models can be fine-tuned on domain-specific datasets, enabling rapid adaptation to new robotic domains, operational environments, or system architectures. This reduces manual feature engineering and accelerates intelligent test optimization. Zheng et al. \cite{zheng2024testing} demonstrate that fine-tuning LLMs on heterogeneous data advances formal verification for autonomous systems. DiaVio \cite{Lu2024DiaVioLD} further supports this by using domain-specific language (DSL) to align natural language descriptions with simulation testing, showcasing how fine-tuned models enhance diagnostic efficiency and accuracy.


Foundation models can aid human testers by providing semantic summaries of test executions, highlighting anomalies, and suggesting critical tests for manual review. This synergy between automated analysis and expert judgment improves robustness in complex or safety-critical scenarios. ScenLaBe \cite{Li2024LargeLM} integrates LLMs with Behavior Trees to generate test scenarios with high behavioral coverage from natural language descriptions of traffic accidents. By leveraging the strengths of LLMs in natural language understanding and Behavior Trees in task decomposition, ScenLaBe efficiently transforms textual descriptions into executable test scenarios, enhancing scenario diversity and complexity. This method not only improves the realism and validity of test cases but also provides valuable support for human-in-the-loop validation.

Overall, the integration of multi-source foundation models into the testing workflow promises to significantly enhance the semantic understanding, adaptability, and efficiency of test optimization for autonomous systems. However, challenges such as model interpretability, computational cost, and domain-specific fine-tuning requirements must be carefully addressed to fully realize the potential of these models in ROSAS testing.



\subsubsection{Neurosymbolic Distillation for Fine-grained Violation Detection}
Further development of neurosymbolic methods is essential, emphasizing scalable and automatic extraction of formal specifications, improved interpretability, and symbolic-guided test optimization. Research should specifically investigate human-in-the-loop refinement, hybrid symbolic-neural validation frameworks, and robust semantic rule integration.

To capture subtle yet critical behavioral deviations such as a small animal crossing during an autonomous landing sequence \cite{liang2025garl}, traditional frame-level coverage or vector similarity is often insufficient. Specifically, as discussed in \ref{frame-to-vector}, purely vectorization-based methods, such as our proposed frame-to-vector approach, inherently struggle with capturing minor but crucial features. These nuanced violations, essential for system safety and reliability, are frequently overlooked or inadequately represented during the vectorization process.

To address this challenge, we propose adopting Neurosymbolic Approaches. Neurosymbolic methods integrate neural network-based learning with symbolic logic, combining the interpretability and structured reasoning capabilities of formal logic with neural model expressiveness \cite{Zheng2025NeuroStrataHN}. By defining explicit symbolic rules with human-in-the-loop feedback, neurosymbolic frameworks can precisely prioritize critical yet subtle dimensions (such as dynamic object violations), enhancing test selection and prioritization processes. Additionally, by increasing the weighting of distance similarity metrics, this approach also improves test suite reduction efficacy, capturing subtle distinctions between similar scenarios.


Neurosymbolic frameworks can embed domain-specific safety rules and semantic constraints directly into the testing process. By prioritizing dimensions related to critical behaviors such as dynamic object detection, safe distance maintenance, or violation of operational design domain boundaries, test optimization can focus on exposing rare but safety-critical faults. This approach ensures that testing efforts are concentrated on the most critical aspects of system behavior, rather than being diffused across less relevant dimensions. Lu et al. \cite{Lu2024SurveyingNA} have conducted an in-depth investigation into the application of neurosymbolic methods, particularly focusing on their potential to address the testability, verifiability, and interpretability of deep learning models. These characteristics are crucial for detecting minor violations in ROSAS testing. Their study highlights how neurosymbolic approaches can effectively combine the powerful data processing capabilities of neural networks with the logical reasoning abilities of symbolic AI, enabling more accurate identification and interpretation of anomalous behaviors including those subtle violations that may otherwise go undetected.



Rather than relying solely on heuristic or search-based prioritization, neurosymbolic methods can utilize formal specifications or learned logical rules to guide test selection and reduction. This transition from search-driven to specification-driven optimization increases reliability, explainability, and alignment with formal safety standards. Recent advancements underscore the effectiveness of neurosymbolic distillation, a technique where formal specifications are extracted from neural policies, facilitating a transition from search-based to specification-based testing, which inherently provides more formal guarantees \cite{blazek2024automated}. Neurosymbolic distillation combines the strengths of neural perception and symbolic reasoning to prioritize and verify fine-grained, semantically meaningful dimensions. For example, the NUDGE framework introduced by Delfosse et al. \cite{delfosse2023interpretable} demonstrates how neurally guided symbolic abstraction can distill neural network behaviors into weighted symbolic logic rules, enabling interpretable and explainable policies. Additionally, Marconato et al.'s BEARS framework \cite{Marconato2024BEARSMN} enhances neurosymbolic models' awareness of reasoning shortcuts, reducing misjudgments of tiny violations as normal behavior. BEARS operates efficiently even with sparse supervision, making it suitable for resource-constrained ROSAS environments. This approach ensures that test selection is not only efficient but also formally grounded, enhancing the robustness of the testing process.


Neural models can extract rich latent representations from sensor data, while symbolic components can distill these into interpretable and verifiable properties. This enables testing frameworks to identify subtle deviations that are meaningful in a safety or functional context, even if they are difficult to detect through raw output comparisons. Singireddy et al. \cite{singireddy2023automaton} propose ``Automaton Distillation", which transfers learned knowledge from deep neural policies into deterministic finite automata (DFAs), representing policies as formal logical structures. By mapping neural Q-values into symbolic transitions within these automata, this approach creates a structured, interpretable, and formally analyzable representation of complex decision processes. The result is an abstraction layer that allows systematic reasoning and validation through formal methods, significantly enhancing robustness and reliability, particularly in complex sequential decision-making scenarios characteristic of ROSAS. Similarly, NS3D \cite{Hsu2023NS3DNG} decomposes complex 3D scene understanding tasks into interpretable symbolic programs and neural modules, improving data efficiency and generalization. In ROSAS testing, this modular approach enables precise semantic understanding of object relationships, facilitating the detection of minor deviations during task execution.


Neurosymbolic systems support iterative human feedback by allowing domain experts to refine or extend symbolic rules based on observed system behaviors. This enables continuous improvement of the test optimization process and ensures that evolving operational knowledge is incorporated into validation efforts. By leveraging human expertise to refine symbolic rules, the testing process can adapt to new challenges and emerging behaviors, maintaining its relevance and effectiveness over time. This idea is further supported by the Recover \cite{Cornelio2024RecoverAN} framework, which integrates LLMs with symbolic reasoning to achieve online detection and recovery of minor violations during robot task execution, such as objects falling or paths being blocked. This approach not only accurately identifies violations but also reduces the hallucination phenomenon of LLMs through symbolic knowledge guidance, thereby generating more reliable recovery plans. Moreover, the Recover framework has great potential for application in ROSAS environments, especially in complex tasks and dynamic settings, where it can significantly enhance the efficiency and reliability of autonomous system testing. Experimental validation in the AI2Thor simulator has demonstrated Recover's superior performance in minor violation detection and recovery, providing an efficient, low-cost, and interpretable neurosymbolic method for ROSAS testing.


By focusing on semantic, task-relevant, and safety-critical aspects of behavior, neurosymbolic approaches can reduce dependence on raw structural or coverage-driven metrics, which often miss subtle but important behavioral nuances in autonomous systems. Instead of relying solely on coverage metrics, neurosymbolic methods provide a more nuanced and context-aware approach to test optimization, ensuring that critical behavioral deviations are detected even if they do not manifest in obvious structural coverage gaps. The PhysORD framework \cite{Zhao2024PhysORDAN} exemplifies this paradigm. It demonstrates high-precision prediction capabilities for vehicle motion in complex off-road environments by integrating physical laws (such as the Euler-Lagrange equations) with neural networks. This method, which combines physical knowledge with data-driven learning, not only enhances the model's generalization ability and data efficiency but also provides the possibility for detecting minor deviations. In ROSAS testing, detecting minor deviations is a crucial aspect of ensuring system safety and reliability. The neurosymbolic framework of PhysORD can precisely model physical processes through its symbolic model while capturing subtle changes in data using neural networks, thereby detecting minor deviations that might be overlooked by traditional methods during testing. For instance, in the testing of autonomous vehicles, PhysORD can be used to detect minor trajectory deviations or dynamic anomalies when vehicles travel over complex terrain, which may be caused by sensor noise, terrain changes, or minor disturbances in vehicle dynamics.

By developing neurosymbolic test optimization frameworks along these lines, future research can significantly improve the ability to detect tiny yet vital behavioral violations, enhancing both the safety and robustness of ROS-based autonomous systems under real-world operational conditions.

\section{Conclusion} \label{sec:conclusion}
In this paper, we conducted the first systematic and comprehensive survey on regression testing optimization techniques specifically tailored for ROS-based autonomous systems. Recognizing that conventional testing approaches from traditional software engineering are inadequate for addressing the unique characteristics of autonomous systems such as dynamic and non-deterministic behaviors, multi-modal sensor integration, asynchronous distributed architectures, and stringent real-time and safety-critical constraints, this survey critically analyzed the applicability, effectiveness, and limitations of existing test optimization methods.

Through a systematic review of 122 representative studies, we categorized test optimization techniques into test case prioritization, test suite minimization, and test case selection. To clearly illustrate the relationships and distinctions among these techniques, a structured taxonomy was proposed, facilitating a deeper understanding of the optimization landscape within autonomous system testing contexts.

Key limitations and gaps of current optimization techniques were identified, including challenges related to semantic adequacy, multi-modal data fusion, tiny violation detection, and oracle deficiencies. To address these limitations, we proposed research insights and outlined promising future research directions. Notably, the development of frame-to-vector coverage metrics, integration of multi-source foundation models, and application of neurosymbolic reasoning techniques emerged as pivotal pathways toward advancing test optimization strategies for ROS-based autonomous systems.

Overall, this survey contributes foundational knowledge and practical guidance for researchers and practitioners alike, laying out a clear and actionable roadmap for future investigations. By systematically bridging existing methodological gaps and highlighting new research opportunities, we aim to foster further advancements toward more reliable, scalable, and safety-assured testing practices in autonomous systems.



\bibliography{sn-article}

\end{document}